\documentclass[preprint, 5p, twocolumn, authoryear]{elsarticle}

\usepackage[pdftex]{hyperref}
\usepackage{graphicx} 
\usepackage{epstopdf} 
\usepackage{dcolumn}
\usepackage{amssymb}
\usepackage{amsmath}
\usepackage{fixltx2e} 
\usepackage{pict2e} 

\begin{document}

\begin{frontmatter}

\title{Measuring social dynamics in a massive multiplayer online game}

\author[cosy]{Michael Szell}
\author[cosy,santafe]{Stefan Thurner\corref{cor1}}
\ead{thurner@univie.ac.at}
\ead[url]{http://complex-systems.meduniwien.ac.at}
\cortext[cor1]{Corresponding author.}
\address[cosy]{Complex Systems Research Group, Medical University of Vienna, Austria}
\address[santafe]{Santa Fe Institute, Santa Fe, NM 87501, USA}

\begin{abstract}
Quantification of human group-behavior has so far defied an empirical, falsifiable approach. 
This is due to tremendous difficulties in data acquisition of social systems.  
Massive multiplayer online games (MMOG) provide a fascinating new way of observing hundreds of thousands of simultaneously socially interacting individuals engaged in virtual economic activities. 
We have compiled a data set consisting of practically all actions of all players over a period of three years from a MMOG played by 300,000 people. This large-scale data set of a  socio-economic unit contains all social and economic data from a single and coherent source. Players have  to generate a virtual income through economic activities to `survive' and are typically engaged  in a multitude of social activities offered within the game. Our analysis of high-frequency log files focuses on three types of social networks, and tests a series of social-dynamics hypotheses. In particular we study the structure and dynamics of friend-, enemy- and communication networks. We find striking differences in topological structure between positive (friend) and negative (enemy) tie networks. All networks confirm the recently observed phenomenon of network densification. 
We propose two approximate social laws in communication networks, the first expressing betweenness centrality  as the inverse square of the overlap, the second relating communication strength to the cube of the overlap.  These empirical laws provide strong quantitative evidence for the Weak ties hypothesis of Granovetter. Further, the analysis of triad significance profiles validates well-established assertions from social balance theory. We find  overrepresentation (underrepresentation) of complete (incomplete) triads in networks of positive ties, and vice versa for networks of negative ties. Empirical transition probabilities between triad classes provide  evidence for triadic closure with extraordinarily high precision. For the first time we provide empirical results for large-scale networks of negative social ties. Whenever possible we compare our findings with data from non-virtual human groups and provide further evidence that online game communities serve as a valid model for a wide class of human societies. With this setup we demonstrate the feasibility for establishing a `socio-economic laboratory' which allows to operate at levels of precision approaching those of the natural sciences.\\[5pt]
All data used in this study is fully anonymized; the authors have the written consent to publish from the legal department of the Medical University of Vienna. 
\end{abstract}

\begin{keyword}
Social network analysis \sep Network theory \sep Massive multiplayer online game \sep Social balance \sep Triadic closure \sep Quantitative social science
\end{keyword}

\end{frontmatter}


\section{Introduction}
\label{sec:introduction}
Quantification of collective human behavior or social dynamics poses a unique, century old challenge. 
It is remarkable to some extent that  mankind knows more about dynamics of subatomic particles than it knows about the dynamics of human groups. The reason for this situation is that the establishment of a fully experimental and falsifiable social science of group dynamics is tremendously complicated by two factors:  First, unlike many problems in the natural sciences, dynamics of societies constitute a \emph{complex system}, characterized by strong and long-range interactions, which are in general not treatable by traditional mathematical methods and physical concepts. Second, data is of comparably poor availability and quality \citep{watts2007tfc,lazer2009css}. Evidently it is much harder to obtain data from social systems than from repeatable experiments on (non-complex) physical systems. Despite these severe problems, it is nevertheless paramount to arrive at a better understanding of collective human behavior. Only recently it became most evident in the context of economics and finance, which costs are associated to misconceptions of human collective behavior. If the dynamics behind collective behavior are going to remain as poorly understood as they are today, without being able to generate statements with predictive value, any attempts of managing crises will turn out not a whit better than illusionary. 

Many complex systems cannot be understood without  their \emph{surroundings}, contexts or boundaries, together with the interactions between these boundaries and the system itself. This is obviously necessary for measuring large-scale dynamics of human groups. Regarding data acquisition it is therefore essential not only to record decisions of individual humans but also the simultaneous state of their surroundings. Further, in any data-driven science the observed system should not be significantly perturbed through the act of measurement. In social science experiments subjects usually are fully aware of being observed  -- a fact that might strongly influence their behavior. Finally, data acquisition in the social sciences becomes especially tiresome on group levels, see e.g. the tremendous efforts which have been undertaken in a classic experiment by \cite{newcomb1961ap}, in which a group of 17 students was congregated, observed, and questioned over several weeks. Traditional methods of social science such as interviews and questionnaires do not only need a lot of time and resources to deliver statistically meaningful assertions, but may introduce well-known biases \citep{carrington2005mam}. To many it might seem clear that social sciences can not overcome these problems, and that therefore social sciences would always remain on a lower quantitative and qualitative level than the natural sciences. 

Both issues, the availability of data, and the possibility to take simultaneous measurements on subjects and their surroundings, might appear in a radically more positive light when looking at massive multiplayer online games (MMOGs) \citep{castronova2005swb}. Such computer games not only allow to conduct \emph{complete} measurements of socially interacting humans, they also provide data at rates comparable to physical experiments. Remarkably, one of the largest collective human activities on the planet is the playing of online games. Currently more than a hundred million people worldwide play MMOGs -- the well-known game \verb|World of Warcraft| alone has more than ten million subscribers as of today. MMOGs exhibit such an enormous success due to offering their players possibilities to experience alternative or second lifes, not only providing (virtual) \emph{economic} opportunities, but also a huge variety of possible \emph{social interactions} among players. Many MMOGs provide rich virtual environments facilitating socialization and interactions on group levels \citep{yee2006mpo, yee2006dma, castronova2005swb}. Motivation of players to participate in MMOGs are highly heterogeneous, ranging from establishing friendships, gain of respect and status within the virtual society, to the fun of destroying the hard work of other players. Besides economical and social interactions, modern MMOGs also offer a component of exploration, e.g. players can explore their `physical' environment, such as specific features of their universe, `biological' details of  space-monsters, etc., and share their findings within `specialist' communities.

From a scientific point of view online games provide a tool for understanding collective human phenomena and social dynamics on an entirely different scale \citep{bainbridge2007srp, castronova2006orv}. In these games \emph{all} information about \emph{all} actions taken by \emph{all} players can be easily recorded and stored in log-files at practically no cost. This quantity of data has been unthinkable in the traditional social sciences where sample sizes often do not exceed several dozens of questionnaires, school classes or students in behavioral experiments. In MMOGs on the other hand, the number of subjects can reach several hundred  thousands, with millions of recorded actions. These actions of individual players are known in conjunction with their surroundings, i.e. the circumstances under which particular actions or decisions were taken. This offers the unique opportunity to study a complex social system: conditions under which individuals take decisions can in principle be controlled, the specific outcomes of decisions can be measured. In this respect social science is on the verge of becoming a fully experimental science \citep{lazer2009css} which should increasingly become capable of making a great number of repeatable and eventually falsifiable statements about collective human behavior, both in a social and economical context.

Another advantage over traditional ways of data acquisition in the social sciences is that players of  MMOGs do not consciously notice the measurement process.\footnote{Players are informed that data is logged for scientific purposes and give their consent, usually  prior to their participation in the game.} These `social experiments' practically do not perturb or influence the sample. Moreover MMOGs not only open ways to explore sociological questions, but -- if economic aspects are part of the game (as it is in many MMOGs) -- also to study economical behavior of groups. Here again economical actions and decisions can be monitored for a huge number of individual players within their social and economical contexts. This means that MMOGs offer a natural environment to conduct \emph{behavioral economics} experiments, which have been of great interest in numerous small-scale surveys, see e.g. \cite{gachter1999cas,henrich2005emc}. It becomes possible to study the \emph{socio-economic unit} of large online game societies.

In the past years we have recorded practically all actions of all players taken in the self-developed, proprietary MMOG \verb|Pardus| which is online since 2004. \verb|Pardus| is an open-ended game with a worldwide player base of more than 300,000 people. Players reside and act within a virtual, persistent futuristic universe and make up their own goals.  Most players  invent and develop their virtual social lifes without constraints by the game setup. The game's environmental topology is given but can be manipulated by the players to some extent. Players self-organize within groups and subgroups, claim territories, decide to go to war, etc., completely on their own accounts. Players typically participate in the game for several weeks to several years. 

Players of \verb|Pardus| characteristically engage in various economic activities to increase their wealth (non-convertible game money): There are numerous possibilities for jobs, such as mining and processing basic resources from the environment, trade, production, assembly  and consumption of commodities, etc. Economic life is embedded in a production tree which provides a basic framework for player-created industries. Trade occurs following simple `rules' within dynamic and demand-oriented virtual markets constituted  by groups of players. Social life within \verb|Pardus| is based on means of communication with fellow players in various forms, such as chat, forum, private messages, which allow  the establishment of e.g. friendships or hostile relations. There are a number of ways  to publicly display one's `status' within the virtual society: Purchase of expensive status symbols, such as space ships, earning of medals of honor for war efforts or for defeating outlaws, etc. These possibilities are not only well used, but constitute an important psychological driving force for many players.

Given the complete data set from the \verb|Pardus|  game, one can identify four major directions of research.

\textbf{1. Network analysis.} It is possible to directly access the dynamics of several types of social networks such as dynamics of friend networks, networks of enemies, or communication networks. Especially the latter offer a fantastic way to directly relate findings in the game with real-world communication networks, such as a data set of cell phone calls which has been recently analyzed from a network perspective \citep{onnela2007als,lambiotte2008gdm}. While there exists some insight into real-world friend networks in the literature, e.g. of the \verb|Facebook| community \citep{golder2007rsi}, there is practically no knowledge of topology and dynamics of enemy networks \citep{labianca2006esl}. Since the time resolution of our data is accurate to one second, it becomes possible to study time courses of global network properties. This way it can be understood  if and how communities show aging effects, such as \emph{densification}, i.e. shrinking diameters and growing average degrees. This phenomenon has been observed in societies and online communities \citep{leskovec2007ged, leskovec2008mes}, as well as in the evolution of scientific fields and cities \citep{bettencourt2007gis,bettencourt2008pme}. The vast majority of social network studies analyze single or at best small numbers of network snapshots. Important exceptions include time resolved studies of an internet dating community \citep{holme2004sat}, the analysis of a university email network \citep{kossinets2006eae}, of the web of scientific coauthorships \citep{ravasz2004eha, newman2001ssc, newman2004cnp}, as well as several large-scale networks of various types \citep{leskovec2007ged}. 

Network growth and re-linking processes  can be directly studied and compared to well-known models, such as e.g. the preferential attachment model \citep{barabasi1999esr} or static relinking models \citep{thurner2007umn}. Preferential attachment dynamics of real-world networks have been verified in a  few recent studies \citep{csardi2007mik,leskovec2008mes,jeong2003mpa}.

\textbf{2. Testing traditional social-dynamics hypotheses.} The  \verb|Pardus| data allows for direct empirical testing of long-standing hypotheses on social network dynamics, such as the \emph{Hypothesis of triadic closure} \citep{rapoport1953sit,granovetter1973swt}, the \emph{Weak ties hypothesis} \citep{granovetter1973swt}, or the \emph{Hypothesis of social balance} \citep{harary1953nbs,doreian1996pas}.

 For quantification purposes we employ network measures such as \emph{betweenness centrality} \citep{freeman1977smc} and \emph{overlap} which measures how often a given pair of nodes has links to other common nodes \citep{onnela2007als}. To our knowledge, no longitudinal measurements of large-scale signed networks exist as of today. One well-known social network study on monks in a monastery can be found in the classic literature \citep{sampson1968npc}, as well as a modern long-time survey of social dynamics in classrooms \citep{jordan2009cit}. These are first attempts of systematic social balance experiments, however being far from conclusive due to limited data and small scales ($10$ to $100$ nodes), and a low number of samples (about $\approx\,10$ observations). Further, the extent of \emph{reciprocity} and \emph{Triad significance profiles} \citep{milo2004sea} together with their dynamics can be directly accessed from the game data. For the quantification of these concepts we use recent technology developed in the context of \emph{motif distributions}. To understand microscopic changes in social network dynamics, transition rates between dyadic and triadic structures can be measured -- yielding parameters needed e.g. for calibrating agent-based models of social network dynamics, as e.g. in \cite{antal2006sbn}. So far these transition rates could only be assumed by model builders and have never been measured in actual societies.

\textbf{3. Economic analysis.} The fact that all players are engaged in economic activities allows for statistically significant measurements of \emph{wealth and income distributions} which can be compared to real economies \citep{yakovenko2009esm,dragulescu2001eap,chatterjee2007ein}. The process of price formation -- and more generally preference relations -- for all goods and services in the game can be observed within the social and economical context of players. All prices of all goods in the game are recorded with a time resolution of one second and can be analyzed with respect to `stylized' facts in real prices, in a straightforward fashion. These `stylized' facts, such as volatility clustering, fat-tailed return distributions, squared auto-correlation decays, etc., are known in detail for traded goods in the real world \citep{cont2001epa}. Further, co-evolution dynamics, i.e. the evolution of economic properties of players (e.g. wealth) as a function of their local social networks, and vice versa, their social evolution as a function of their economic network, can be extracted from the data. The theoretical literature on co-evolving networks is relatively sparse \citep{biely2005psd,biely2009sed}; to our knowledge there exist practically no measurements on this issue so far.

\textbf{4. Group formation and dynamics -- gender and country aspects.} Players have the possibility to create and join `alliances' (communities) in the game, which allow them to streamline ideas, join forces for common projects, or coordinate aims and believes. The dynamics, formation, interaction, and disappearance of these cohesive communities can be investigated readily in the data. Players choose to have a male or female character allowing all data to be partitioned into female and male networks. This offers the possibility to search for behavioral differences in `networking' patterns between female and male players. Further, economic productivity and communication patterns can be analyzed gender-specifically. The same holds for country specifics of players.

In this work -- the first of a series --  we focus on the first two of the above directions: analysis of complex network structures/dynamics and testing sociological hypotheses. Along these lines we establish further evidence that online game communities may serve as a model for real world communities. It is not obvious {\em a priori} that a population of online players is a representative sample of real-world societies \citep{williams2009vcr}. However, several recent studies are providing evidence that human behavior on a collective level is remarkably robust, meaning that statistical differences of real-world communities and game-societies are often marginal \citep{johnson2009hgf, jiang2009ooa}. 

The paper is organized as follows. In section \ref{sec:thegamestudied} we present the game, describe the sample of players and explain their modes of communication. We introduce the  three types of  social networks studied. Section \ref{sec:networks} contains the network measures used in our analyses. Results are presented in section \ref{sec:results}  and  are discussed in section \ref{sec:discussion}. Finally we conclude in section \ref{sec:conclusion}.


\section{The game}
\label{sec:thegamestudied}
\subsection{Overview}
\verb|Pardus| (\href{http://www.pardus.at}{\ttfamily{http://www.pardus.at}}) is a browser-based MMOG in a science-fiction setting, open to the public and played since September 2004. A browser-based MMOG is characterized by a substantial number of users playing together in the same virtual environment connected by an internet browser. For a detailed categorization of online games see \cite{bartle2004dvw,castronova2005swb}.

In \verb|Pardus| every player onws an account with one \emph{character} per game universe; players are forbidden to operate multiple accounts. A character is a pilot owning a spacecraft with a certain cargo capacity, roaming the virtual universe trading commodities, socializing, and much more, `to gain wealth and fame in space' (\href{http://www.pardus.at/index.php?section=about}{\ttfamily{http://www.pardus.at/index.php?section=about}}). The main component of \verb|Pardus| consists of trade simulation with a society of players heavily driven by social factors such as friendship, cooperation or competition. There is no explicit `winning' in \verb|Pardus| as there is no inherent set of goals nor allowed or forbidden `moves' (with a few exceptions mainly concerning decent language and behavior towards fellow players). \verb|Pardus| is a \emph{virtual world} or \emph{synthetic world} with a gameplay based on socializing and role-playing, with  interaction of player characters with others  and with non-player characters as its core elements \citep{castronova2005swb}.

\begin{figure}[tb]
    \begin{center}
        \includegraphics{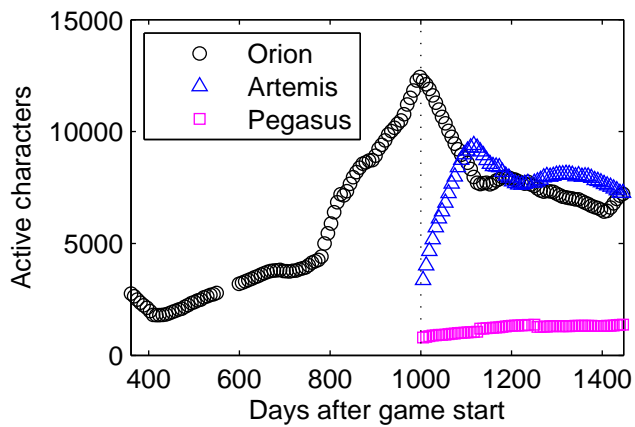}    
    \end{center}
    \caption{Evolution of number of active characters in the game universes. The large increase of players in Orion between days $\approx\,800$ and 1,000 is due to ad campaigns after  October 2006. At day 1,000 (dotted line) Artemis and Pegasus were opened. Some thousand players abandoned their Orion characters focusing on their new Artemis characters. This  explains the mirrored development in these two universes after day 1,000.}
    \label{fig:activecharacters}
\end{figure}

There exist three separate game universes: Orion, Artemis, and Pegasus. Presently \verb|Pardus| is actively played by $\approx\,$14,000 players, over 300,000 have registered so far. Orion and Artemis each inhabit $\approx\,$6,500 active characters, Pegasus with its $\approx\,$1,400 characters is for  paying customers only. 
We count existing characters who have mastered the game's tutorial environment as active (characters which are inactive for 120 days get deleted automatically, see section \ref{sec:playercensus}). 
Figure \ref{fig:activecharacters} depicts the evolution of universe populations for the time range where data is available (see section \ref{sec:dataanalyzed}). The majority plays the game for free, paying members receive  \emph{Premium accounts} which bestow them with additional features not available to users with \emph{Free account} status (such as the possibility of character creation in the Pegasus game universe). 
Orion was opened on September 14$^{\mathrm{th}}$ 2004, Artemis and Pegasus 1,000 days later, on June 10$^{\mathrm{th}}$ 2007.
Between universes it is impossible to move, trade, or exchange game money.
The universes are independent.\footnote{This is not entirely correct since some players have openly revealed their identities, i.e. they have disclosed which characters they are controlling in different game universes. It is not clear how many attempts have been made to copy existing social ties between universes. Although it is discouraged, it may happen that e.g. vendettas between players who are aware of their mutual identities in different universes are carried out within more than one universe. We discuss this issue in more detail in section \ref{sec:discussion}.}

\subsection{The data analyzed}
\label{sec:dataanalyzed}
Daily database backups recorded at 05:32 GMT\footnote{This time was chosen for the daily backup and maintenance scheduler because it is the time of lowest player activity.} are available from 2005-09-09 to 2008-09-01. The day 2005-09-09 is the 360$^{\mathrm{th}}$ after 2004-09-14, i.e. the 360$^{\mathrm{th}}$ day after Orion was opened. Backups from the following dates are not available due to unknown reasons: 2006-03-24 to 2006-04-23, 2006-04-28 to 2006-04-30, 2006-10-24 to 2006-10-26, 2007-03-20, 2007-05-10, 2007-09-21, 2008-02-09, 2008-06-09. Since we have complete data only for Artemis and Pegasus (with the exception of three days), and because Artemis has more active characters than Pegasus, in this work we refer only to  Artemis data. Results are remarkably robust between universes. For clarity only weekly data points are shown in all time evolution plots, except for figs.~\ref{fig:linksversusnodes} and \ref{fig:tspevolution}.

\subsection{Players and census of characters}
\label{sec:playercensus}

\subsubsection*{Age and nationality}
\begin{figure}[htb]
    \begin{center}
        \includegraphics{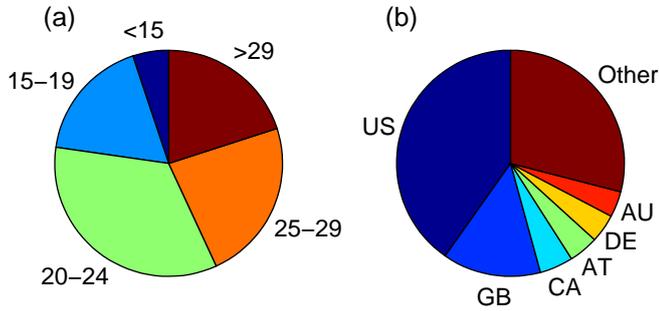}    
    \end{center}
    \caption{Distribution of player age (a) , nationality (b).}
    \label{fig:agecountry}
\end{figure}

In a poll taken in the \verb|Pardus| forums in the beginning of 2005 the age of players was assessed. 
From a total of 255 votes, 5\% reported their age to be  less than 15 years, 18\% between 15--19, 34\% between  20--24,  23\% between  25--29, and  20\% are older than 29 years, fig.~\ref{fig:agecountry} (a). The distribution of player nationalities can be estimated by technical means and reads approximately as follows: United States (US) 40\%, United Kingdom (GB) 14\%, Canada (CA) 5\%, Austria (AT) 4\%, Germany (DE) 4\%, Australia (AU) 4\%, Other 29\%, fig.~\ref{fig:agecountry} (b).

\subsubsection*{Lifetimes of characters}

\begin{figure}[tb]
    \begin{center}
        \includegraphics{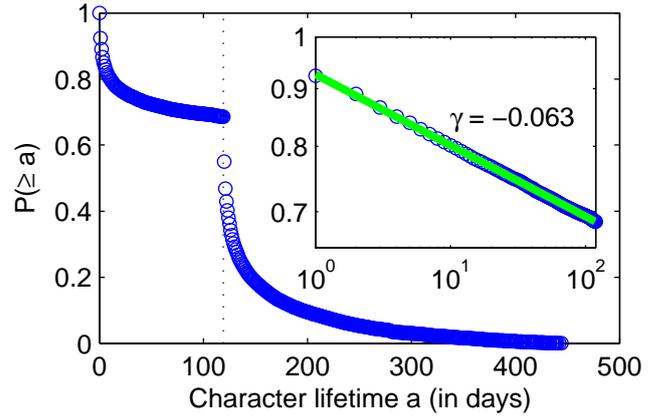}    
    \end{center}
    \caption{Cumulative distribution of character lifetimes (in days) of all 16,980 characters who existed at
    least for one day but not on the last one. The dotted line marks day 120. The inset shows the distribution of lifetimes of $<120$ days in a double-logarithmic scale.}
    \label{fig:lifetimes}
\end{figure}

Characters are automatically deleted after an inactivity (not logging in) period  of 120 days. Additionally, every player has the option to delete her account or characters at any time. Rarely, it happens that accounts get deleted due to breaking of game rules, such as the operation of multiple accounts. We call all deletions which are not due to inactivity \emph{self-induced}. Figure \ref{fig:lifetimes} shows the cumulative distribution of character lifetimes (in days) of all 16,980 characters who existed for at least one day, but not on the last one. If a character's lifetime lies before day 120 (dotted line), her deletion could have been self-induced only. If a character's lifetime is longer than 120 days, her deletion was either self-induced or automatic due to inactivity. These two regimes are evident in the figure: the regime of only self-induced deletion follows a power-law with exponent $\gamma = -0.063$. 
In the second regime, the distribution is neither a power-law nor an exponential probably due to the overlay  of the two deletion schemes.

Of all characters $7.6\%$ have a lifetime of 0 days, i.e. they delete themselves on the first day of their existence. At least $31.4\%$ of all deletions are self-induced, and $\approx\,13\%$ of all characters become inactive after  their first day. 

\subsubsection*{Gender of characters}
When signing up for the first time, players have to choose between a male and female character; this decision is irrevocable.
Depending on gender a male or female \emph{avatar} (profile-like picture) of characters is displayed in certain places in the game. In the Artemis universe, $\approx 90\%$ of all characters are male.  

\subsection{Structure of the universe}
\begin{figure}[t]
    \begin{center}
        \includegraphics[width=0.452\textwidth]{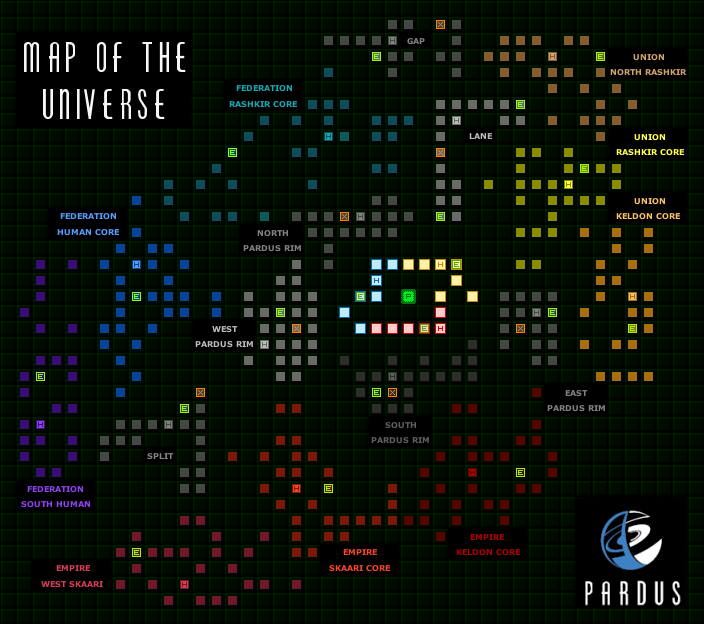}    
    \end{center}
    \caption{Map of a universe. Each colored square represents a sector, consisting of 15$\times$15 fields on average. Colors indicate cluster membership. Nearby sectors are connected by wormholes (not shown).}
    \label{fig:pardusuniverse}
\end{figure}

\begin{figure}[t]
    \begin{center}
        \includegraphics[width=0.4\textwidth]{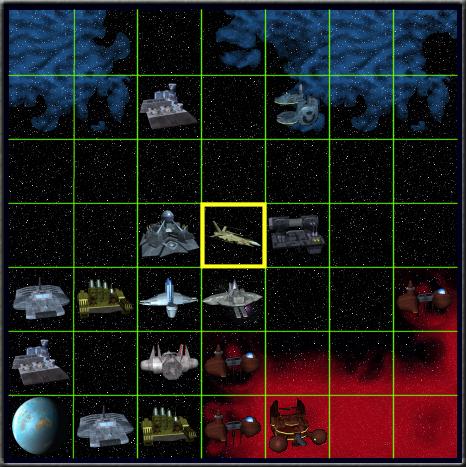}    
    \end{center}
    \caption{Space chart. A cut-out of 7$\times$7 fields of the universe (green lines, not shown in-game) visible to every player on his navigation screen; the current position is in the central field (yellow box). By clicking on another field, the ship moves to that location and all fields are updated so the ship is again located in the center.}
    \label{fig:spacechart}
\end{figure}

Space in \verb|Pardus| is two-dimensional. Each game universe is divided into 400 \emph{sectors}, fig.~\ref{fig:pardusuniverse}, each sector consisting of 15$\times$15 \emph{fields} on average. Fields are the smallest units of space and are displayed as 64$\times$64-pixel images in-game. They form a square grid on which continuous ship movement is possible by clicking on the desired destination field within the \emph{space chart}. This chart is a 7$\times$7 fields cut-out of the universe visible to every player with their current position located on the central field, see fig.~\ref{fig:spacechart}. A sector's boundary is impenetrable; moving between nearby sectors is possible by tunneling through field objects called \emph{wormholes}. A collection of nearby $\approx\,$20 sectors is called a \emph{cluster}. The typical spatial range of activities of a character is usually confined to one cluster for several weeks or longer.

\subsection{Action Points -- the unit of time}
Every game action carried out by a  player (trade, travel, etc.) costs certain amount of so-called \emph{Action Points} (APs). These points can not exceed a maximum of 6,100 APs per character. For characters owning less APs than their maximum, every six minutes 24 APs are automatically regenerated, i.e. 5,760 APs per day. Once a player's character is out of APs, she has to wait for being able to play on. As a result the typical \verb|Pardus| player logs in once a day to spend all her APs on several activities within a few minutes (for each character/universe). This makes APs, the game's unit of time, the most valuable factor: Those players who use their APs most efficiently can experience the fastest progress or earn the highest profits. Social activities such as chatting (see section \ref{sub:socialinteraction}) do not consume APs. Highly involved players usually spend a lot more real time on the game's features of socialization as well as on planning and coordinating their future moves than on actually spending their APs.

\subsection{Trade and industry}
The currency of game money within \verb|Pardus| is the so-called \emph{credit}. This money is not convertible to real money. Every player starts her life with 5,000 credits. Since most assets needed for making progress -- such as ships, ship equipment, buildings -- are traded in credits, it is of basic interest to earn money during the game (the richest players currently posess hundreds of millions of credits). There exist a number of possibilities to do this, usually through participation in the economy.

There exist over 30 commodity types. Some of these grow in the environment and can be mined, such as gas from nebulas or ore from asteroids. Most commodities however are processed from less valuable (more basic) commodities in player-owned firms. For example, a brewery manufactures expensive liquor out of cheap energy, water, food, and chemical supplies. Every player has the possibility to construct a small number of such buildings. Since production chains follow a fixed production tree, coordination of several players is needed to build up a profitable \emph{industry}. Most end-products, i.e. products at the top of the production tree (which cannot be reused as upkeep in other buildings), are usable commodities. For example, manufactured drugs may be consumed to gain APs, droid modules can be installed for powerful building defenses against hostile attacks, etc. Therefore, tangible needs generated by the society  play a driving role for motivating the development of industries.

Besides player-created firms there exist bases owned by the game system, which trade and regularly consume most commodity types. Prices there are exclusively determined by local supply and demand: when commodities are available in abundance, prices are low, when only a few are available, prices rise.

Through these mechanics players find themselves strongly encouraged to take part in the struggles of economic life, as known from the real world: collaboration, competition, cartelization, fraud, etc. We will analyze the \verb|Pardus| economy in detail in a separate work.

\subsection{Social interaction}
\label{sub:socialinteraction}
There are three ways for players to communicate inside the game, facilitating social activity. Players may use these facilities independently from game-mechanic states (such as their ship's location within the universe, their wealth, etc.):
\begin{enumerate}
    \item \emph{Chat:} \verb|Pardus| offers several built-in chat channels per universe where players can simultaneously communicate with many others. Chat entries scroll up and disappear; thus the chat is well-suited only for temporary talks.
    \item \emph{Forum:} In the forum, messages, called \emph{posts}, can consist of several lines and stay for a long time. This enables more thorough discussions than in chat.
    Posts are organized within \emph{threads} which correspond to a topic. There are universe-specific subforums as well as global subforums which can be accessed from all universes.
    \item \emph{Private message:} Within a universe it is possible to send private messages (PMs) to any other player; this action and the PM's content is only seen by sender and receiver -- a system similar to email. When a PM is sent the receiver gets immediately notified in a status bar. PMs always have exactly one recipient. Presently a daily total of $\approx\,10,000$ PMs are exchanged within \verb|Pardus|. 
 \end{enumerate}

\subsection{Friends and enemies}
For a small amount of APs, players can mark others as \emph{friend} or \emph{enemy}.
This can be done for  any reason. The marked characters are added to the markers personal \emph{friends} or \emph{enemies list}. Additionally, every player has a personal \emph{friend of} and \emph{enemy of} list, displaying all players who have marked them as friend or enemy, respectively. When being marked or unmarked as friend or enemy the affected player immediately receives an informatory system message. It is only possible to mark someone either as friend or as enemy, but not both. 

We stress that friend/enemy lists and friend of/enemy of lists are \emph{completely private}, meaning that no one except the marking and marked players have information about ties between them. It is not possible to see second degree neighbors (e.g. friends of friends) or the number of ties another player has.\footnote{On 2008-08-24 the \emph{profile} feature of the game was extended, allowing players with Premium accounts to publicly display their numbers of friends or enemies. Since this feature was introduced at the very end of our last measured data (2008-09-01) and only a negligible proportion of players are making use of it, it is irrelevant here.}
Note that this is in contrast to many online social networking services such as \verb|Facebook|, where usually second degree neighbors and number of friends are visible.
Thereby \verb|Pardus|' system does not introduce potentially strong biases concerning accumulation of friends (some users may tend to accumulate friends for the main purpose of increasing their publicly visible number of friends \citep{golder2007rsi}). Our data thus represents a more realistic social situation, in the sense that social ties are not immediately accessible but need to be found out by e.g. communication with or careful observation of others.

\begin{figure}[tb]
    \begin{center}
        \mbox{(a)}\hspace*{-0.5cm}\includegraphics[width=0.49\textwidth]{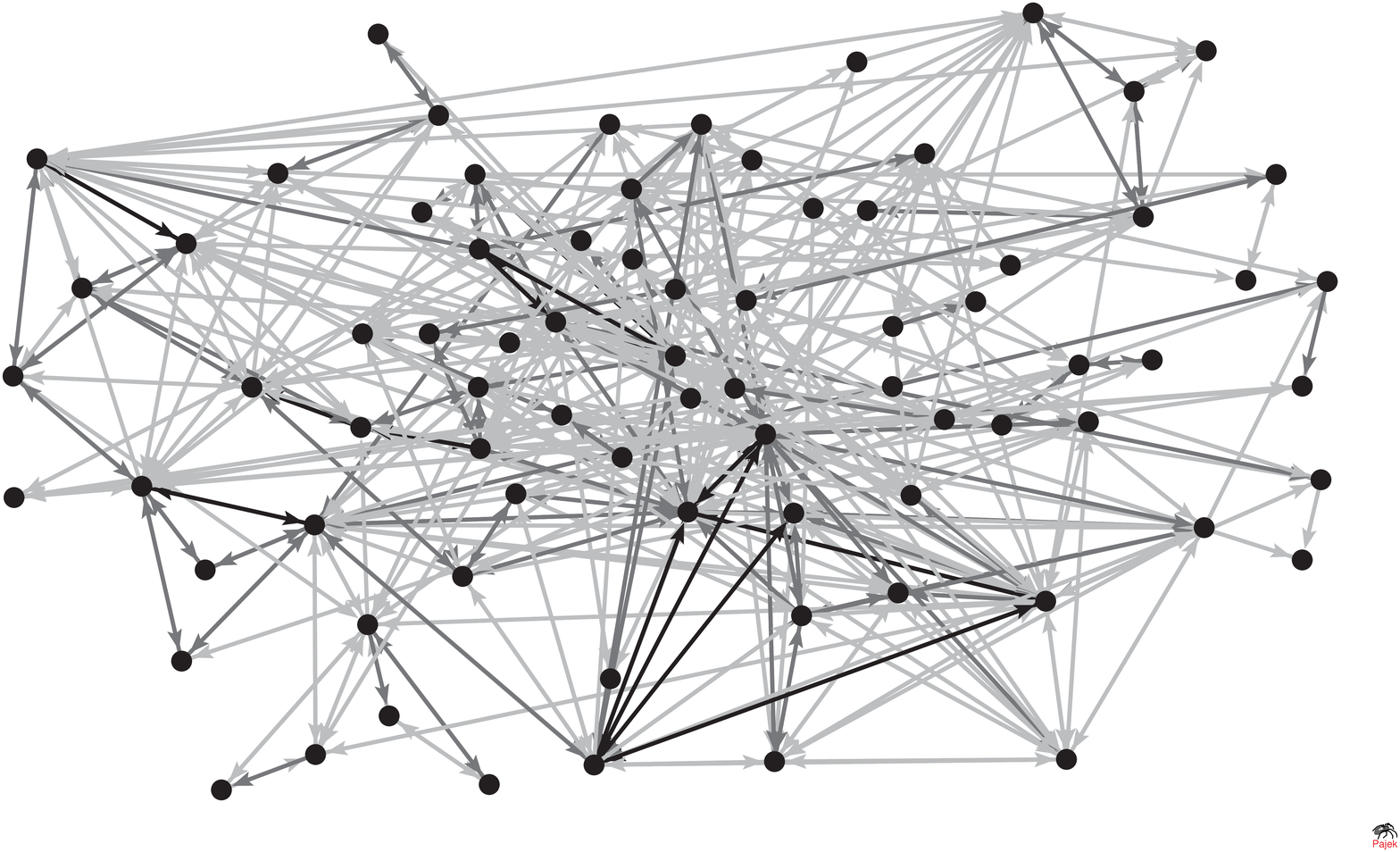}
        \mbox{(b)}\hspace*{-0.5cm}\includegraphics[width=0.49\textwidth]{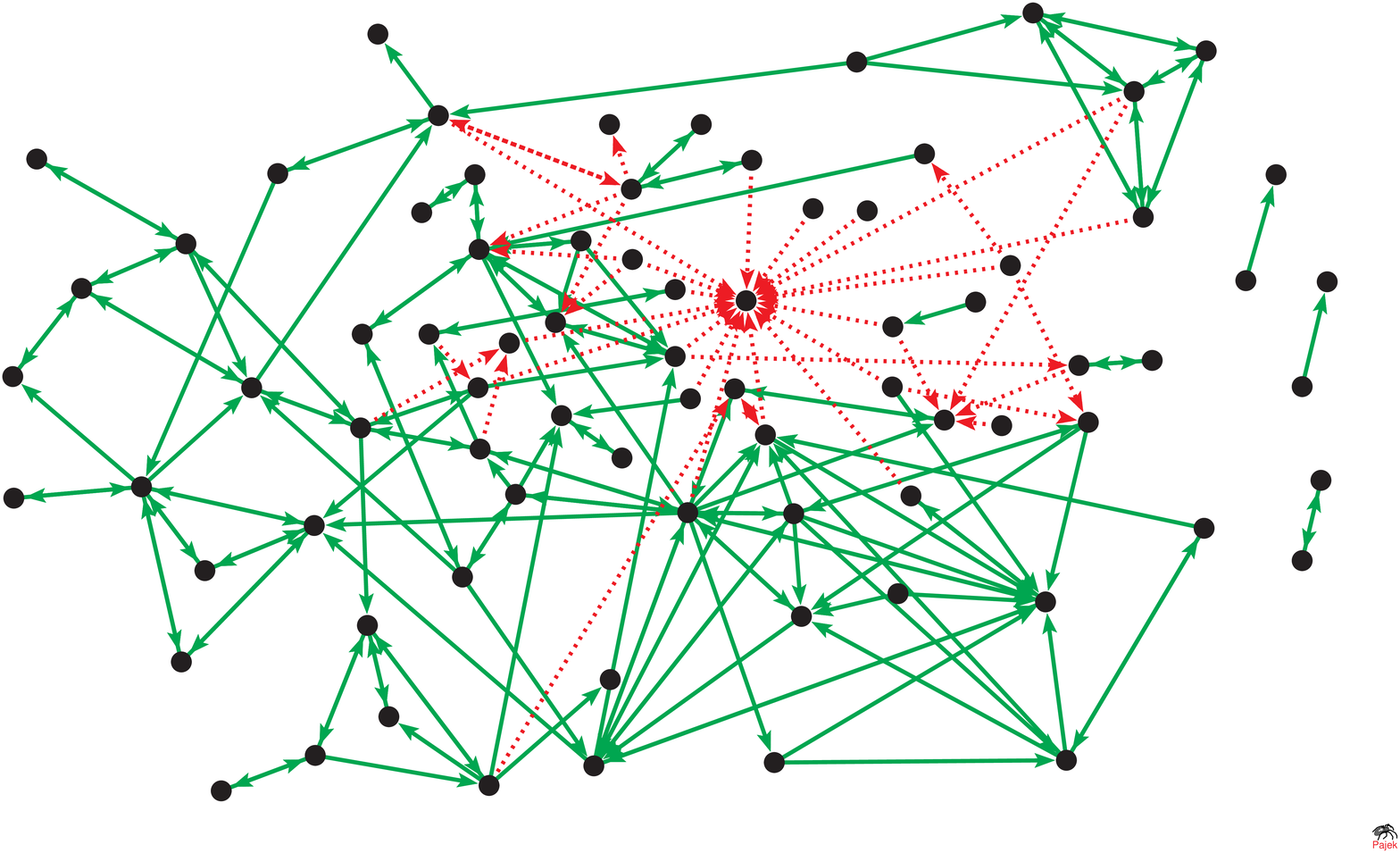}
    \end{center}
    \caption{(a) Accumulated PM communications over all 445 days between 78 randomly selected individuals who existed on the first and last day. Link colors of light gray, gray, and black correspond to 1--10, 11--100 and 101--1000 PMs sent, respectively. 
        (b) Friend (green, solid) and enemy (red, dashed) relations on day 445 between the same individuals. See our Youtube channel \href{http://www.youtube.com/user/complexsystemsvienna}{\ttfamily{http://www.youtube.com/user/complexsystemsvienna}} for animated time evolutions of these networks.
    }
    \label{fig:pajeknet}
\end{figure}

Besides character names and online status being displayed on every player's personal PM contacts page for quick access, the friends and enemies lists serve game-mechanic purposes: friends/enemies are automatically or optionally included/excluded for certain actions. For example, enemies of building owners are not able to use the services offered in the respective places. Note that  friend and enemy markings need not necessarily denote \emph{affective} friendships or enmity, they rather indicate a certain degree of cooperative or uncooperative stance motivated by affective and/or cognitive incentives. However, we assume these two motives to coincide to a great extent, e.g. it seems highly unlikely that someone marked as enemy/friend due to rational considerations at the same time constitutes the affective opposite of friend/enemy within the game (and vice versa). PMs as well as friend and enemy relations can be displayed as networks, fig.~\ref{fig:pajeknet}, see also section \ref{networkextraction}. 

\subsection{Universe-specific characteristics}
Orion has the longest history but has some missing data at its beginning (see section \ref{sec:dataanalyzed}). Orion has converged to social structures and ties which are relatively constant over time, as seen in a number of measures (not shown). In contrast, Artemis and Pegasus can be observed from their start on, and measuring the evolution of  social networks from  the time of their birth is possible. Here we do not face the problem of the `missing past' \citep{leskovec2007ged}. Further, since character creation in Orion and Artemis is free but not so in Pegasus, the former universes display considerably higher fluctuations in player numbers and activity rates than the latter. 

\subsection{A note on equilibrium  and `steady state'}
The  \verb|Pardus| universes are virtual worlds, permanently evolving without a scheduled end. They are far from being in equilibrium due to constantly added new and changed game features, fluctuations in player numbers, large-scale collective actions of payers,  etc. The distinction between transient phase and steady state may be hard to determine for certain network properties, due to the large amount of strong mutual influences, naturally present in systems of such high complexity.


\section{Networks}
\label{sec:networks}
\subsection{Definitions}
\subsubsection*{Graph}
In mathematical terms, networks are described by \emph{graphs} \citep{wasserman1994sna, dorogovtsev2003enb}. An undirected graph $\mathcal{G} = (\mathcal{N}, \mathcal{L})$ is defined as a pair of sets, the node set $\mathcal{N}$ containing all nodes $n_i$ and the link set $\mathcal{L}$ containing unordered pairs $l_{ij} := \{n_i, n_j\}$ denoting those nodes which are connected by an undirected link (edge). A directed graph (\emph{digraph}) has a link set $\mathcal{L}$ which contains \emph{ordered} pairs $l_{ij} := (n_i, n_j)$ marking nodes which are connected by a directed link (arc) going from $n_i$ to $n_j$. The expressions $N$, $L$ denote cardinalities of the respective sets. A graph is called \emph{complete} if connections between all pairs of nodes exist.

\subsubsection*{Symmetrization}
The \emph{symmetrization} or \emph{reflexive closure} of a digraph $\mathcal{G} = (\mathcal{N}, \mathcal{L})$ is constructed as follows: Start with $\mathcal{G}^* = (\mathcal{N}, \mathcal{L}^*)$, where $\mathcal{L}^*$ is an empty link set, and for all pairs of nodes $n_i$ and $n_j$ add the undirected link $l_{ij}$ to $\mathcal{L}^*$ if the directed link $l_{ij} \in \mathcal{L}$ or if $l_{ji} \in \mathcal{L}$.

\subsubsection*{Weighted graph}
In unweighted graphs all links are treated equally. A \emph{weighted graph} is a generalization in which the weight $w_{ij}$ of a link $l_{ij}$ may take any non-zero real value.

\subsubsection*{Dyad}
A dyad is a (sub)graph consisting of two nodes. A directed dyad can be a \emph{null dyad} (no links), \emph{asymmetric} (one link, going in one direction), or \emph{mutual} (two links, one going in one direction and the other going in the opposite one). 

\subsubsection*{Signed graph}
A \emph{signed digraph} is a pair $(\mathcal{G}, \sigma)$, where $\mathcal{G} = (\mathcal{N}, \mathcal{L})$ is a digraph and $\sigma\mathrm{:}\,\mathcal{L}\rightarrow\{-1, +1\}$ is a sign function assigning each directed link a binary value, e.g. in the context of social networks denoting positive or negative relationship \citep{doreian1996pas}. We write $s_{ij}$ short for $\sigma(l_{ij})$, and set $s_{ij} := 0$ when $l_{ij}$ does not exist.

Every signed digraph has a valency matrix $V$ with entries $v_{ij}$ defined as \citep{harary1965smi}:
\begin{align}
	v_{ij} & = o \qquad \mathrm{if}\,s_{ij} = s_{ji} = 0\nonumber \\ 
	v_{ij} & = p \qquad \mathrm{if}\,s_{ij} + s_{ji} > 0\nonumber \\
	v_{ij} & = n \qquad \mathrm{if}\,s_{ij} + s_{ji} < 0\nonumber \\
	v_{ij} & = a \qquad \mathrm{otherwise}
\end{align}
These entries correspond to null  ($o$) dyads, to dyads with only positive ties ($p$) , to dyads with only negative ties ($n$), and to dyads with one positive and one negative tie ($a$ for ambivalent relationship), respectively.

\subsubsection*{Degree}
In an undirected graph the \emph{degree} $k_i$ of a node $n_i$ is the number of links connecting to it. All $k_i$ nodes which are directly linked to $n_i$ are called (nearest) \emph{neighbors} of $n_i$.
A node with degree 0 has no neighbors and is called \emph{isolated}. We denote the average degree of all nodes in a network by $\bar{k}$. In a directed graph the in-degree $k^{\mathrm{in}}_i$ of a node $n_i$ is the number of its incoming links, the out-degree $k^{\mathrm{out}}_i$ the number of its outgoing links. We denote the average degree of all nearest neighbors of a node $n_i$ by $k^{\mathrm{nn}}_i$. We denote the average degree of all nearest neighbors of all nodes as a function of degree $k$ by $k^{\mathrm{nn}}(k)$ .

\subsubsection*{Geodesic}
In an undirected graph, the \emph{geodesic} or \emph{shortest path} $g_{ij}$ of two nodes $n_i$ and $n_j$ is the smallest number of links one needs to get from $n_i$ to $n_j$. 
If a graph is disconnected, i.e. there exist at least two non-empty sets of non-connected nodes (called \emph{components}), geodesics between all nodes of different components are set to $\infty$. The average geodesic of a random graph is $\bar{g}_{\mathrm{r}} \approx \ln N / \ln \bar{k}$ \citep{dorogovtsev2003enb}.

\subsubsection*{Clustering coefficient}
The \emph{clustering coefficient} $C_i$ of node $n_i$ in an undirected graph is the ratio between the number $y_i$ of links between its $k_i$ neighbors and the number of all possible links $k_i(k_i - 1)/2$ between them,
\begin{equation}
    C_{i} := \frac{2y_i}{k_i(k_i - 1)}.
    \label{eq:clusteringcoefficient}
\end{equation}
The network's clustering coefficient $C$ is the average over all clustering coefficients, $C = (1/N) \sum_i C_i$. A random graph's clustering coefficient $C_{\mathrm{r}}$ is given by $C_{\mathrm{r}} = \bar{k} / N$ \citep{dorogovtsev2003enb}.

\subsubsection*{Efficiency}
Global efficiency of an unweighted network $\mathcal{G}$ with $N$ nodes is defined as 
\begin{equation}
    E_{\mathrm{glob}}(\mathcal{G}) := \frac{2}{N(N - 1)} \sum_{i \neq j \in \{1,\ldots,N\}}{g_{ij}^{-1}}.
    \label{eq:efficiencyglobal}
\end{equation}
Global efficiency $E_{\mathrm{glob}}$ can be thought of  as a measure how efficiently information 
is exchanged over a network, given that all nodes are communicating with all other nodes concurrently.
Local efficiency $E_{\mathrm{loc}}$, as a measure of a system's fault tolerance is defined as
\begin{equation}
    E_{\mathrm{loc}}(\mathcal{G}) := \frac{1}{N} \sum_{i \in \{1,\ldots,N\}}{E_{\mathrm{glob}}}(\mathcal{G}_i),
    \label{eq:efficiencylocal}
\end{equation}
where $\mathcal{G}_i$ is the graph of all neighbors of node $n_i$ (not containing $n_i$).
Both values $E_{\mathrm{glob}}$ and $E_{\mathrm{loc}}$ are in the interval $[0,1]$. Note that global efficiency is a reasonable approximation for the inverse geodesic in unweighted graphs; local efficiency is a good approximation for the clustering coefficient when most local networks $\mathcal{G}_i$ are not sparse \citep{latora2001ebs}.

\subsubsection*{Reciprocity}
\emph{Reciprocity} measures the tendency of individuals to reciprocate connections, i.e. the creation of mutual instead of asymmetric dyads \citep{wasserman1994sna}. Following \cite{holme2004sat}, a naive reciprocity index can be defined by
\begin{equation}
    R := \frac{L}{L^*}-1,
    \label{eq:reciprocitynaive}
\end{equation}
where $L^*$ is the number of undirected links in the reflexive closure of the digraph.\footnote{The factor 2 of equation (1) in \citep{holme2004sat} is dropped since we identify pairs of directed links of mutual dyads with single undirected links in the construction of the reflexive closure.} Values of $R = 0$ and $R = 1$ stand for no mutual dyads and mutual dyads only, respectively. Reciprocity may also be quantified by defining the fraction 
\begin{equation}
    r^{*} := \frac{L^{\leftrightarrow}}{L},
    \label{eq:reciprocityproblematic}
\end{equation}
where $L^{\leftrightarrow} \equiv 2(L - L^*)$ counts the number of directed links in all mutual dyads of the digraph. Due to conceptual problems with $r^{*}$, \cite{garlaschelli2004plr} we use the following reciprocity index 
\begin{equation}
    \rho := \frac{r^{*}-\bar{a}}{1-\bar{a}} \in [\rho_{\mathrm{min}}, 1],
    \label{eq:reciprocitygarlaschelli}
\end{equation}
with $\bar{a} := \frac{L}{N(N-1)}$ measuring the ratio of observed to possible directed links, and $\rho_{\mathrm{min}} := -\frac{\bar{a}}{1-\bar{a}} \in [-1, 0]$ for $\bar{a} \leq 1/2$ (the expression $\rho_{\mathrm{min}}$ makes sense for $\bar{a} \leq 1/2$. Otherwise it is not possible to have $L^{\leftrightarrow} = 0$). The index $\rho$ allows to distinguish between \emph{reciprocal} ($\rho > 0$), \emph{areciprocal} ($\rho = 0$) and \emph{antireciprocal} ($\rho < 0$) networks. Further, $\rho$ enables a clear ordering of networks independent of link density which is not possible with $r^{*}$ \citep{garlaschelli2004plr}.

\subsubsection*{Assortativity}
Assortative mixing coefficients are the Pearson correlation coefficients of the degrees at either ends of a link \citep{newman2002amn}:
\begin{equation}
    r = \frac{\overline{ k_{\mathrm{to}}k_{\mathrm{from}} } - \overline{ k_{\mathrm{to}} } \overline{ k_{\mathrm{from}} }}{\sqrt{\overline{ k_{\mathrm{to}}^2 }-\overline{ k_{\mathrm{to}}}^2} \sqrt{\overline{ k_{\mathrm{from}}^2 }-\overline{ k_{\mathrm{from}} }^2}} \in [-1, 1] .
    \label{eq:assortativemixingcoefficient}
\end{equation}
Bars denote averages, $k_{\mathrm{to}}$ and $k_{\mathrm{from}}$ index the (in-, out- or undirected) degrees of nodes  at the beginning and end of links, respectively. Following \cite{holme2004sat} we measure assortativity $r_{\mathrm{undir}}$ in the reflexive closures as well as coefficients for all four combinations of in- and out degrees in the directed networks: $r_{\mathrm{inin}}$, $r_{\mathrm{inout}}$, $r_{\mathrm{outin}}$, $r_{\mathrm{outout}}$. A positive degree--degree correlation coefficient indicates assortativity, i.e. the tendency of nodes with high (low) degrees connecting to nodes with high (low) degrees, a negative correlation means  disassortativity, i.e. the tendency of nodes with high (low) degrees connecting to nodes with low (high) degrees.

\subsubsection*{Bridge}
A \emph{bridge} is a link which, when removed, increases the amount of disconnected components in the graph by one \citep{wasserman1994sna}. A link is a \emph{local bridge} of degree $i$ if its removal causes its endpoints to have geodesic $i$ \citep{granovetter1973swt}.

\subsubsection*{Overlap}
Overlap of two neighboring nodes measures the amount of neighbors common to both of them. We adopt the definition used in \cite{onnela2007als},
\begin{equation}
    O_{ij} := \frac{m_{ij}}{(k_i-1)+(k_j-1)-m_{ij}} \in [0, 1],
    \label{eq:overlap}
\end{equation}
where $m_{ij}$ is the number of neighbors common to both nodes $n_i$ and $n_j$. A value of $O_{ij}=0 \,\, (1)$  corresponds to an empty (identical) common neighborhood of nodes $n_i$ and $n_j$. 

\subsubsection*{Betweenness}
Link betweenness centrality, short \emph{link betweenness} or \emph{load}, is defined for an undirected link $l_{ij}$ by 
\begin{equation}
    b_{ij} := \sum_{n_e \in V} \sum_{n_f \in V\setminus{\left\{n_e\right\}}} \frac{\theta_{ef}(l_{ij})}{\theta_{ef}},
    \label{eq:edgebetweennesscentrality}
\end{equation} 
where $\theta_{ef}(l_{ij})$ is the number of geodesics between $n_e$ and $n_f$ that contain $l_{ij}$, and $\theta_{ef}$ is the total amount of geodesics between $n_e$ and $n_f$ \citep{onnela2007als}. Betweenness can be viewed as a measure of \emph{traffic} if e.g. all pairs of nodes exchange information at the same rate \citep{dorogovtsev2003enb}.

\subsubsection*{Largest connected component}
In graphs with infinitely many nodes one observes the emergence of a \emph{giant component} when crossing a percolation threshold \citep{dorogovtsev2003enb}. The emerging giant component is the only component holding infinitely many nodes. In finite graphs, we call the component having the highest number of nodes \emph{largest connected component}. We denote the fraction of nodes being in the largest connected component by $\Gamma$.

\begin{figure}[tb]
    \begin{center}
        \input{szell2009pardussoc1images/triadclasses.pic}
    \end{center}
    \caption{The 16 isomorphism classes of triads and their ids.}
    \label{fig:triadclasses}
\end{figure}

\subsubsection*{Triad}
A \emph{triad} is a (sub)graph consisting of three nodes. In a digraph there exist 16 isomorphism classes of triads \citep{harary1965smi}. We adopt the notation of \cite{milo2002nms} for the 13 connected classes and label the unconnected ones by $a$, $b$ and $c$, see fig.~\ref{fig:triadclasses}. Within the group of connected triad classes, seven are complete.\footnote{We write connected short for weakly connected, i.e. every two nodes are joined by a semipath \citep{harary1965smi}. We write completeness short for completeness of the reflexive closure, i.e. a directed triad is complete if it contains no null dyads, it is incomplete otherwise.}

\subsubsection*{Triad significance profile}
The \emph{triad significance profile} (TSP) is the vector of statistical significances of each connected triad class compared to random networks drawn from the $U(X_{*+}, X_{+*}, M^{*})$ distribution, i.e. of random networks having identical in/out degrees and equally likely numbers of mutual dyads for each node \citep{roberts2000sms,milo2002nms}. Statistical significance of a triad class $i$ is measured by the $Z$ score
\begin{equation}
    Z_i = \frac{(N^{\mathrm{real}}_{i} - \bar{N}^{\mathrm{rand}}_{i})}{\mathrm{std}(N^{\mathrm{rand}}_{i})},
\end{equation}
where $N^{\mathrm{real}}_{i}$ is the frequency of occurence of the triad class in the considered network, and 
$\bar{N}^{\mathrm{rand}}_{i}$ and $\mathrm{std}(N^{\mathrm{rand}}_{i})$ are the average frequency of occurence and the standard deviation in an ensemble of random networks drawn from $U(X_{*+}, X_{+*}, M^{*})$. The TSP is the normalized  vector of all 13 $Z$ scores, 
\begin{equation}
    {\rm TSP}_i = \frac{Z_i}{\left(\sum_{i = 1}^{13} Z_i^2\right)^{1/2}}
\end{equation}
Note that the TSP emphasizes the \emph{relative} significances of triad classes, constituting an appropriate comparison parameter for networks of arbitrary sizes \citep{milo2002nms}.

\subsection{Network extraction}
\label{networkextraction}
We represent all measured networks as digraphs with nodes representing characters. Note that we do not consider isolated nodes, i.e. characters having no PM communication or friend/enemy relations. In the following we use `link' short for `directed link'.

\subsubsection*{Private messages -- communication networks}
The first set of networks is extracted by considering all PM communications on a weekly timescale. Within the timeframes $[d-6, d]$ for all days $d>6$ all PMs between all characters (who exist over these timeframes) were used to define the PM network at day $d$: a weighted link pointing from node $n_{i}$ to node $n_{j}$ is placed if character $i$ has sent at least one PM to character $j$ within a given week. Weights correspond to the total number of PMs sent within this week.  Figure~\ref{fig:pajeknet} (a) illustrates a subgraph of PM networks of accumulated PM communications over all 445 days between 78 randomly selected characters.

\subsubsection*{Friends and enemies}
Friend and enemy markings constitute the second and third sets of extracted networks: A link is placed from $n_{i}$ to $n_{j}$ if character $i$ has marked character $j$ as friend/enemy. Note that friend/enemy markings exist until they are removed by players (or as long as the related players exist), while PM networks are constructed through an accumulating process. Friend and enemy networks are unweighted, since it is not possible mark friends/enemies more than once.

Since links of friend- and enemy networks never coincide (it is not possible to mark someone as both friend and enemy), we can consider the union of friend- and enemy networks as signed networks. Figure~\ref{fig:pajeknet} (b) illustrates a subgraph of the friend/enemy network of day 445. Note the intense cliquishness/reciprocity of friends and the strong enemy \emph{in-hub}. We show below that these features are typical for these networks.


\section{Results}
\label{sec:results}

In this section we measure social networks as constituted by different relations between players, focusing on the \emph{evolution} of network properties. We provide a round-up of the most important results in section \ref{sec:discussion}.

\subsection{Testing preferential attachment}
The model of preferential attachment (PA) asserts that nodes which link to a network for the first time tend to attach to nodes with high degrees, i.e. to `popular' nodes \citep{barabasi1999esr}. In the extracted directed networks of friends (enemies) we assume this popularity (`disdain') being well expressed by the in-degree. We call characters who get connected to a network for the first time \emph{newcomers}. To test whether evolutions of present networks display a PA bias we measure in-degrees of characters who are marked by newcomers as friend (enemy). Whenever there exists a link $l_{ij}$ on day $d+1$ which has not existed on the previous day $d$ we say that a (one-day) \emph{link event} has taken place between $n_i$ and $n_j$ on day $d$; we call $n_i$ the \emph{source} and $n_j$ the \emph{destination} of this event.

\begin{figure}[tb]
    \begin{center}
        \includegraphics{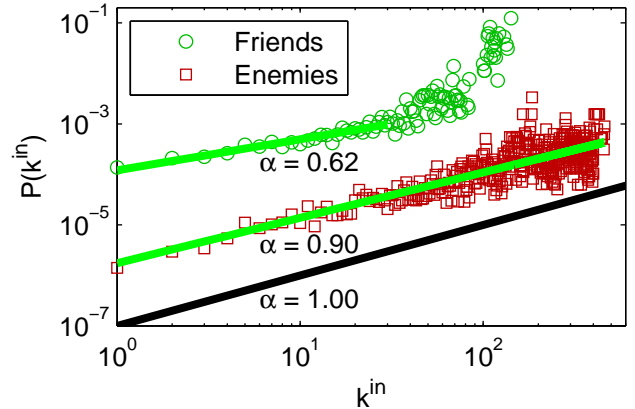}
    \end{center}
    \caption{Empirical probability $P(k^{\mathrm{in}})$ for newcomers connecting to nodes with in-degree $k^{\mathrm{in}}$. Data is used between days 200 and 400. The black line depicts  slope $\alpha = 1$ and indicates the linear dependence assumption needed in the PA model. Green lines denote least squares fits. Values for enemies are vertically displaced  by a factor $0.1$  for better visibility.}
    \label{fig:pa}
\end{figure}

In the classic model of PA it is assumed that the probability $P$ of a newcomer connecting to an existing node $n_i$ with in-degree $k^{\mathrm{in}}_i$ is $P(k^{\mathrm{in}}) \propto \left({k^{\mathrm{in}}}\right)^{\alpha}$ with $\alpha = 1$. Figure \ref{fig:pa} shows $P(k^{\mathrm{in}})$ versus $k^{\mathrm{in}}$ for friend and enemy networks; all  link events between newcomers and their destinations have been used  from day 200 to 400. Least squares fits in double-logarithmic scale yield an exponent of $\alpha = 0.62$ for friend markings with $k^{\mathrm{in}} < 30$, and $\alpha = 0.90$ for all enemy markings. We observe an increased upward bending for players having in-degrees larger than about 100, i.e. for very popular players. These findings are fully consistent with other game universes and other time ranges (not shown).

\subsection{Relations between PM partners}
\begin{figure}[tb]
    \begin{center}
        \includegraphics{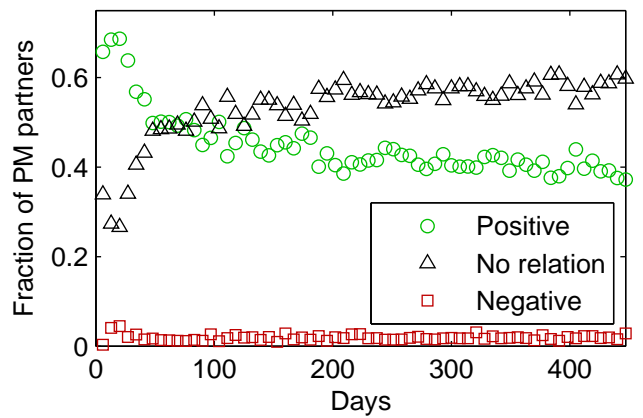}    
    \end{center}
    \caption{Fraction of PM partners per relation class.}
    \label{fig:fractionpmpartners}
\end{figure}

A connection between PM networks and friend/enemy networks can be made visible by partitioning all pairs of characters $\{n_i, n_j\}$ into four classes of friend and/or enemy relations, corresponding to the possible valency matrix entries $v_{ij}$ of a signed digraph. These classes $o, p, n, a$ correspond to dyads without friend/enemy ties ($o$), dyads with asymmetric or mutual friend markings ($p$), dyads with asymmetric or mutual enemy markings ($n$), dyads with one friend and one enemy marking ($a$). Figure \ref{fig:fractionpmpartners} depicts the fraction of all PM partners as partitioned into these classes. Relations of class \emph{a} are not shown since they almost never appear. From the fact that the majority ($> 95\%$) of PM partners consists of positively related characters ($\approx 40\%$ on the last day) and of characters having no friend or enemy relation ($\approx 58\%$ on the last day), we expect a much stronger correlation of PM networks with friend networks than with enemy networks.

\subsection{Measurement of basic network properties}
\begin{figure*}[tb]
    \begin{center}
        \includegraphics{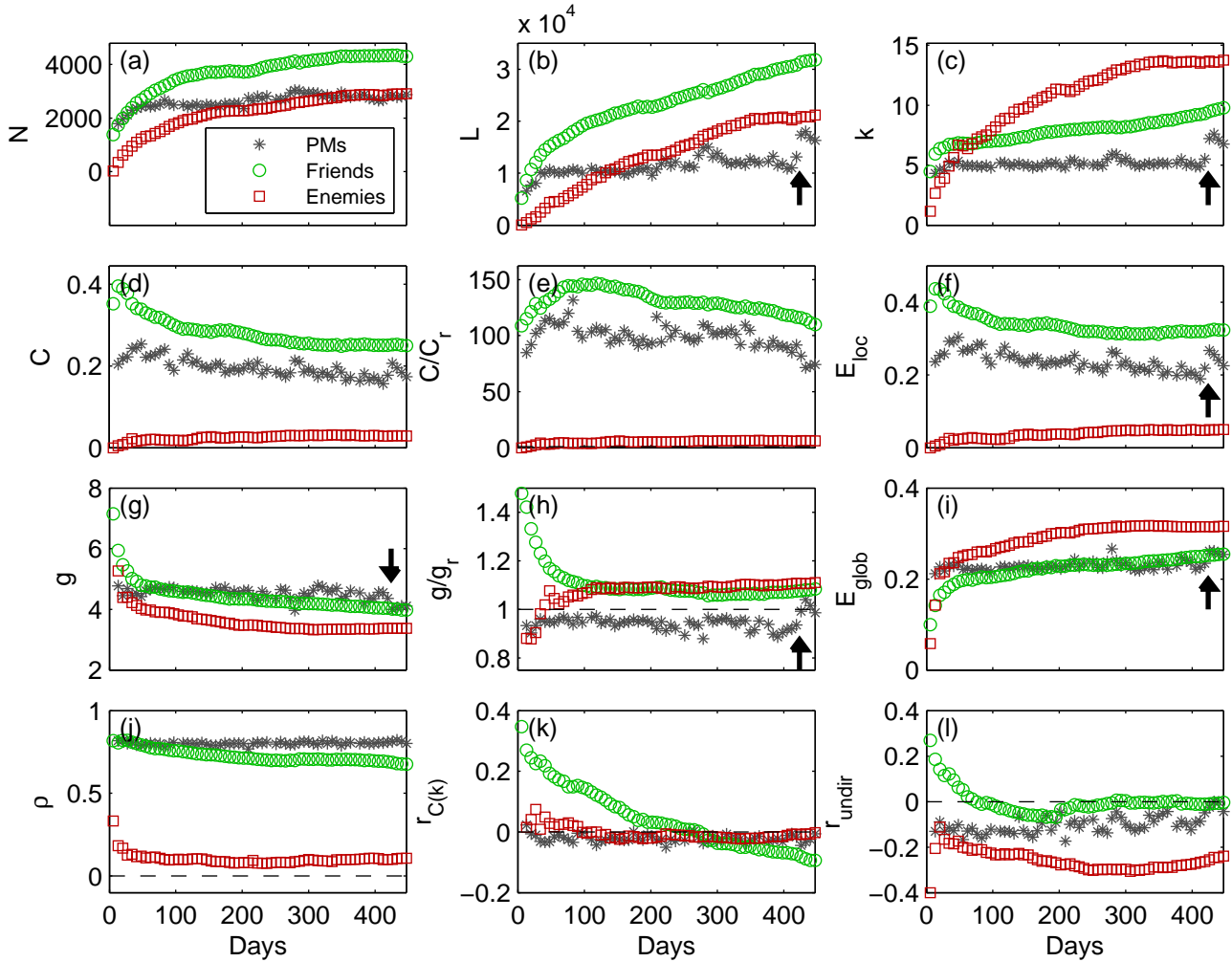}    
    \end{center}
    \caption{Network properties: (a) number of nodes $N$, (b) number of (directed) links $L$, (c) average degree $\bar{k}$, (d) clustering coefficient $C$, (e) clustering coefficient $C$ divided by clustering coefficient of corresponding random graph $C_\mathrm{r}$, (f) local efficiency $E_{\mathrm{loc}}$, (g) average geodesic $\bar{g}$, (h) average geodesic $\bar{g}$ divided by average geodesic of corresponding random graph $\bar{g}_\mathrm{r}$, (i) global efficiency $E_{\mathrm{glob}}$, (j) reciprocity $\rho$, (k) assortative mixing coefficient (of clustering in undirected network) $r_{C(k)}$, (l) assortative mixing coefficient (in undirected network) $r_\mathrm{undir}$. Arrows mark the outbreak  of an in-game war  at day 422.}
    \label{fig:networkproperties}
\end{figure*}

We measure the time evolution of the following basic network properties: number of nodes $N$, directed links $L$ and average degree $\bar{k}$, relative size of largect connected component $\Gamma$, average geodesic $\bar{g}$, clustering coefficient $C$, as well as the comparison values $\bar{g}/\bar{g}_{\mathrm{r}}$ and $C/C_{\mathrm{r}}$. Average degrees, geodesics and clustering coefficients are measured on the reflexive closures of the networks. Geodesics and clustering coefficients were calculated using the \verb|MatlabBGL| package\footnote{We used version 4.0. \\
\href{http://www.stanford.edu/~dgleich/programs/matlab_bgl}{\ttfamily{http://www.stanford.edu/$\sim$dgleich/programs/matlab\_{}bgl}}.}, which efficiently implements standard procedures such as Johnson's algorithm for finding all geodesics in sparse graphs. The measured network properties are displayed in figs.~\ref{fig:networkproperties} to \ref{fig:fractionnodesinlcc}, values of days 50, 150, and 445 are shown in table \ref{tab:networkproperties}. Cumulative distributions of in- and out-degrees of the last day's networks are depicted in figures \ref{fig:pkckknn} (a)--(c).

We make the following observations.

\begin{table*}[tb]
	  \centering
		  \begin{tabular}{ l  c c c | c c c | c c c }
 & \multicolumn{3}{c}{PMs} & \multicolumn{3}{c}{Friends} & \multicolumn{3}{c}{Enemies} \\
 & day 50 & day 150 & day 445 & day 50 & day 150 & day 445 & day 50 & day 150 & day 445 \\
\hline
& & & & & & & & & \\
$N$ & 2,466 & 2,461 & 2,879 & 2,712 & 3,709 & 4,313 & 1,253 & 2,161 & 2,906 \\
$L$ & 10,705 & 9,773 & 16,272 & 15,367 & 21,563 & 31,929 & 4,468 & 11,077 & 21,183 \\
$\bar{k}$ & 5.25 & 4.80 & 6.77 & 6.85 & 7.36 & 9.79 & 6.69 & 9.77 & 13.77 \\
& & & & & & & & & \\
$C$ & 0.24 & 0.19 & 0.17 & 0.34 & 0.28 & 0.25 & 0.02 & 0.03 & 0.03 \\
$C/C_{\mathrm{r}}$ & 112.14 & 99.10 & 74.08 & 133.30 & 143.32 & 109.52 & 3.47 & 5.87 & 6.13 \\
$\bar{g}$ & 4.45 & 4.71 & 4.11 & 4.78 & 4.45 & 3.97 & 3.99 & 3.64 & 3.38 \\
$\bar{g}/\bar{g}_{\mathrm{r}}$ & 0.94 & 0.95 & 0.99 & 1.16 & 1.08 & 1.08 & 1.06 & 1.08 & 1.11 \\
$E_{\mathrm{loc}}$ & 0.29 & 0.23 & 0.23 & 0.38 & 0.34 & 0.32 & 0.02 & 0.04 & 0.05  \\
$E_{\mathrm{glob}}$ & 0.23 & 0.22 & 0.25 & 0.19 & 0.22 & 0.25 & 0.25 & 0.28 & 0.32 \\
& & & & & & & & & \\
$\rho$ & 0.79 & 0.79 & 0.80 & 0.79 & 0.73 & 0.68 & 0.12 & 0.09 & 0.11 \\
$r_{C(k)}$ & -0.03 & -0.02 & -0.01 & 0.19 & 0.08 & -0.09 & 0.02 & -0.02 & -0.00 \\
$r_{\mathrm{undir}}$ & -0.13 & -0.17 & -0.04 & 0.06 & -0.06 & -0.00& -0.19 & -0.23 & -0.24 \\
$\Gamma$ & 0.981 & 0.991 & 0.987 & 0.929 & 0.952 & 0.973 & 0.954 & 0.975 & 0.992\\
		  \end{tabular}
	  \caption{Network properties at days 50, 150, and the last day 445, for PM, friend and enemy networks.}
	  \label{tab:networkproperties}
\end{table*}

\subsubsection*{Growing average degrees, shrinking geodesics}
Average degrees are growing, fig.~\ref{fig:networkproperties} (c). Merely the enemy network reaches a steady state shortly before day 400. Geodesics decrease, fig.~\ref{fig:networkproperties} (g).

\subsubsection*{Links versus nodes}
The $L$ versus $N$ curves for the most part have slopes between 1 and 2, fig.~\ref{fig:linksversusnodes}. Due to heavy fluctuations and limited extension over $N$, fits to power-laws are not reliable.

\subsubsection*{Average geodesics close to random network value}
For all networks the comparison parameter $\bar{g}/\bar{g}_{\mathrm{r}}$ lies well within the band $\left[0.5, 2\right]$ as reported for various scale-free networks by \cite{dorogovtsev2003enb}, fig.~\ref{fig:networkproperties} (h). It fluctuates slightly above 1 for enemy and friend networks, growing for the former, decreasing for the latter. In PM networks $\bar{g}/\bar{g}_{\mathrm{r}}$ is slightly below 1 for most timepoints.

\subsubsection*{Changing clustering coefficients}
Clustering coefficients of friend networks decrease, those of enemy networks increase, fig.~\ref{fig:networkproperties} (d). Concerning $C/C_{\mathrm{r}}$, values fall in friend and PM networks but grow in enemy networks, fig.~\ref{fig:networkproperties} (e). $C/C_{\mathrm{r}}$ curves of PM networks fall between the curves of enemy and friend networks. Decreasing clustering coefficients have been reported for coauthorship networks \citep{ravasz2004eha}, online social networks \citep{hu2009dmo}, and appear in a model of growing social networks \citep{jin2001sgs}. Note that $C/C_{\mathrm{r}}$ has a high value for friend networks ($C/C_{\mathrm{r}} > 100$), as is expected for most social networks of positive ties \citep{newman2003wsn}.

\subsubsection*{Positive reciprocity}
All networks are reciprocal, fig.~\ref{fig:networkproperties} (j).  At the last day the PM network has a reciprocity of $\rho \approx 0.80$, the friend network has $\rho \approx 0.68$ after having reached a maximum of $\rho \approx 0.83$ in the first days. Excluding the first few days, the reciprocity indices of enemy networks lie around $\rho \approx 0.1$. The naive reciprocity $R$ displays qualitatively similar behavior (not shown). Low reciprocities in enemy networks may be explained by deliberate refusal of reciprocation, to demonstrate aversion by lack of any response.

\cite{holme2004sat} report a naive reciprocity value around $R = 0.4$ for a message network, a value much lower than our measured ones around $R \approx 0.7$ in PM networks. We suspect two factors responsible for this discrepancy: A higher community coherence -- i.e. more social pressure to respond -- in \verb|Pardus|, and a possibly high inactivity rate of users on the dating site. Probably for the same reason of community coherence, reciprocities $\rho \approx 0.8$ of \verb|Pardus| PM networks are well above $\rho = 0.194$, a value reported for messages in email networks \cite{garlaschelli2004plr}.

\subsubsection*{No assortativity}
For PM networks, all five considered assortative mixing coefficients reach steady state values slightly below zero, fig.~\ref{fig:networkproperties} (l) (only plots of $r_{\mathrm{undir}}$ are shown). PM networks are therefore disassortative, i.e. a player who sends/receives PMs to/from players with many PM-partners displays a slight tendency of having few PM-partners and vice versa. For a possible explanation see \cite{holme2004sat} who attest this observation as being in contrast to collaboration networks, for which positive assortativity has been measured. There it has been claimed that in friend networks individuals are substitutable and negative mixing is optimal. The approximate steady state of friend networks displays no clear tendency towards assortativity or disassortativity after a transient phase of falling assortativity. Note that by using an assortativity profile it is possible to uncover families of networks, similar to TSPs \citep{milo2004sea,foster2009eds}.

\subsubsection*{Structural change of PM networks due to times of war}
On day 422 a war between a substantial number of players broke out  in the game universe. A structural change of PM networks is identifiable, most clearly in the number of links $L$, average degree $\bar{k}$, average geodesic $\bar{g}$, local efficiency $E_{\mathrm{loc}}$ and global efficiency $E_{\mathrm{glob}}$, see arrows in fig.~\ref{fig:networkproperties}.

\subsubsection*{Growing largest connected component}
\begin{figure}[t]
    \begin{center}
        \includegraphics{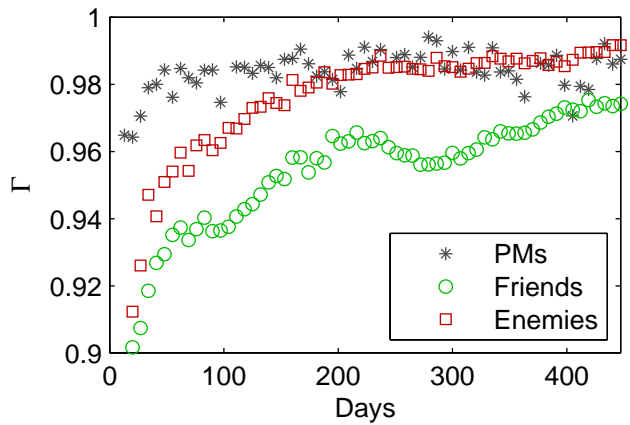}    
    \end{center}
    \caption{Fraction of nodes in the largest connected component $\Gamma$.}
    \label{fig:fractionnodesinlcc}
\end{figure}

The fraction of nodes in the largest connected component, $\Gamma$, is growing in friend and enemy networks over almost all 445 days, fig.~\ref{fig:fractionnodesinlcc}. On the last day we find $\Gamma \approx 0.973$ for friends and  $\Gamma \approx 0.992$ for enemies. The value  for PM networks fluctuates around $\Gamma \approx 0.985$.

\subsection{Overlap versus betweenness and communication strength}
\begin{figure}[!t]
    \begin{center}
        \includegraphics[width=0.49\textwidth]{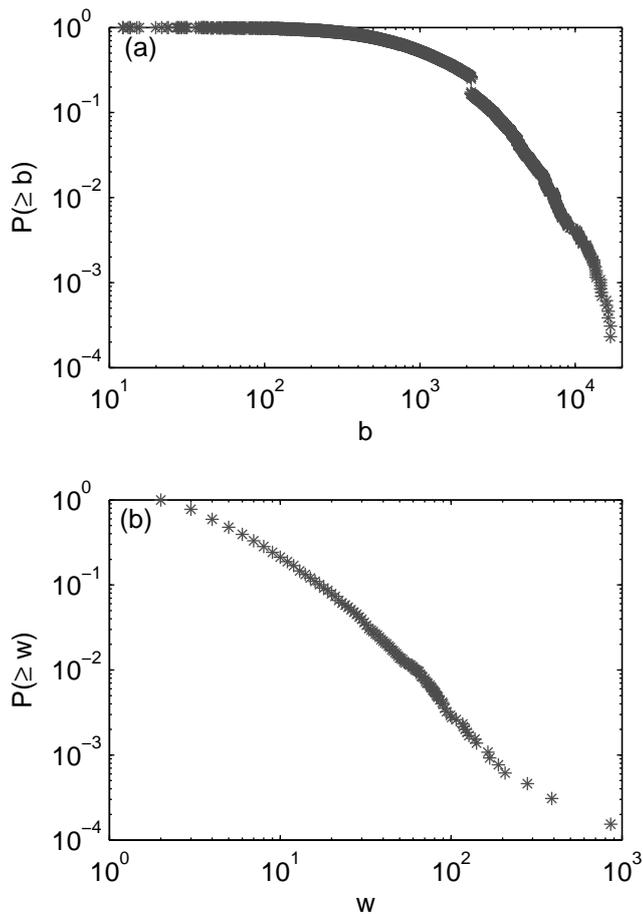}
    \end{center}
    \caption{Cumulative distribution of (a) betweenness and (b) weight values in the mutual part of the largest connected component of the PM network at the last day.}
    \label{fig:pbpw}
\end{figure}

\begin{figure}[!t]
    \begin{center}
        \includegraphics{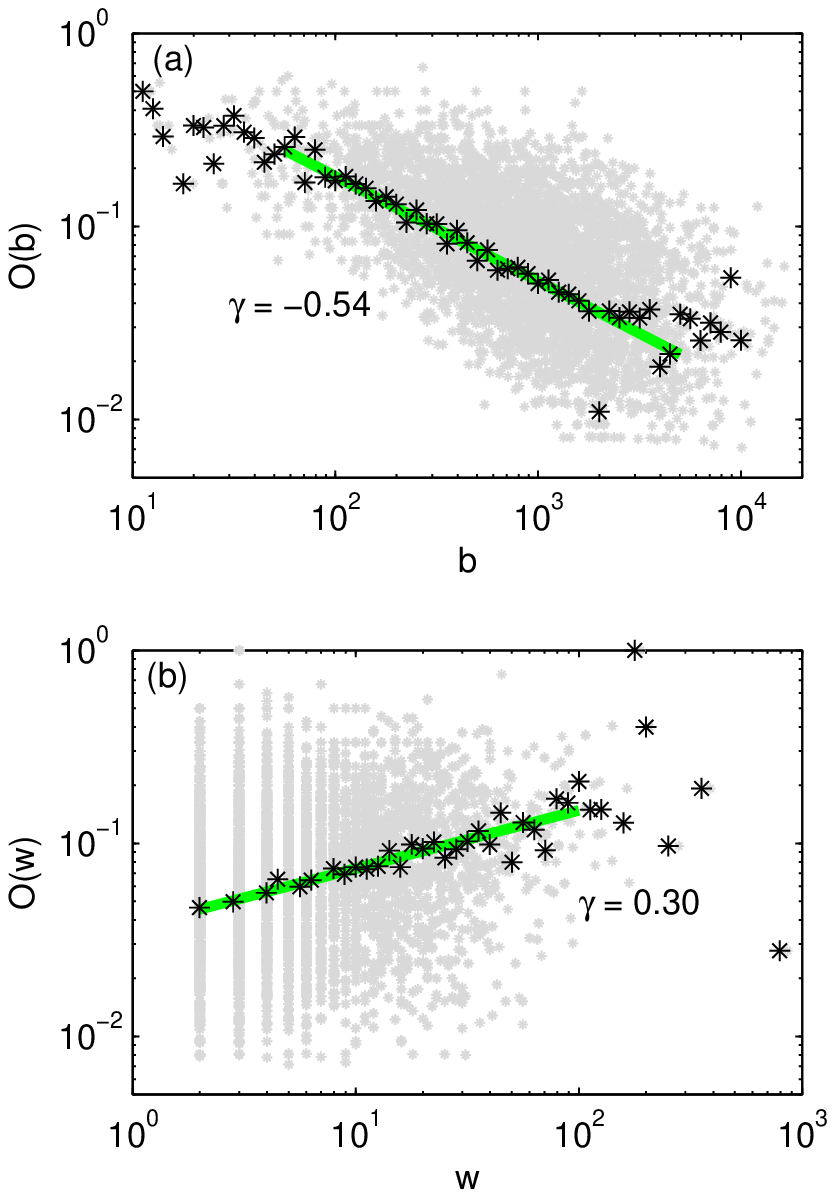}
    \end{center}
    \caption{Overlap versus (a) betweenness and versus (b) weight in the mutual part of the largest connected component  of the PM network at the last day. Light gray markers show individual  overlap values of the links. Note that 2,336 links (out of 6,502) have overlap $O = 0$. Black markers denote logarithmically binned averages, green lines are least squares fits.}
    \label{fig:obow}
\end{figure}

For validating sociological hypotheses (see section \ref{sec:discussion}) we measured overlap versus PM weight and overlap versus betweenness in the largest connected component of the mutual part of the last PM network.
We used this reduced network because it can be directly compared to \cite{onnela2007als}.  Results on {\em full} PM networks and on PM networks accumulated over different time-spans are very similar however (not shown). Betweenness for all links was calculated with the algorithm provided in the \verb|MatlabBGL| package. Results are compiled in  in figs. \ref{fig:obow} and \ref{fig:pbpw}.

\begin{figure}[tb]
    \begin{center}
        \includegraphics{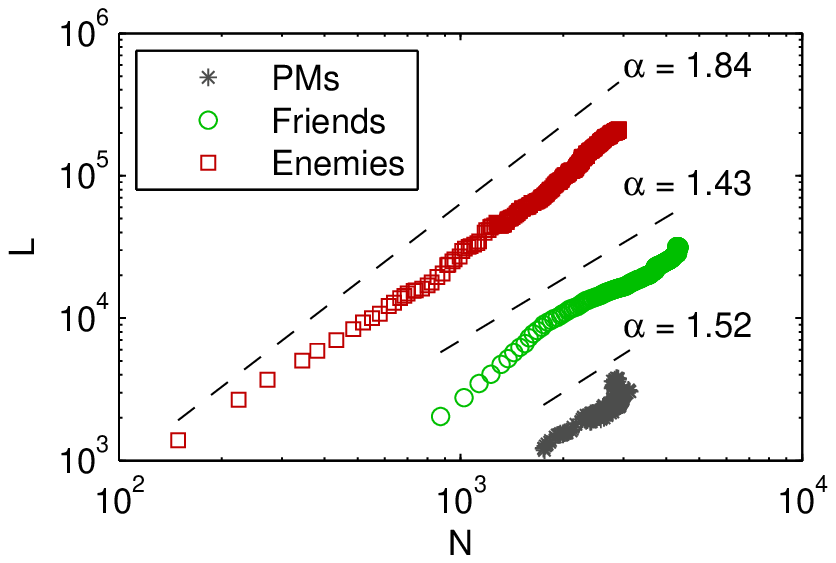}    
    \end{center}
    \caption{Links $L$ versus nodes $N$ for all timepoints. Values for enemies and PMs are shifted vertically for  clarity.}
    \label{fig:linksversusnodes}
\end{figure}

\begin{figure*}[tb]
    \begin{center}
        \includegraphics{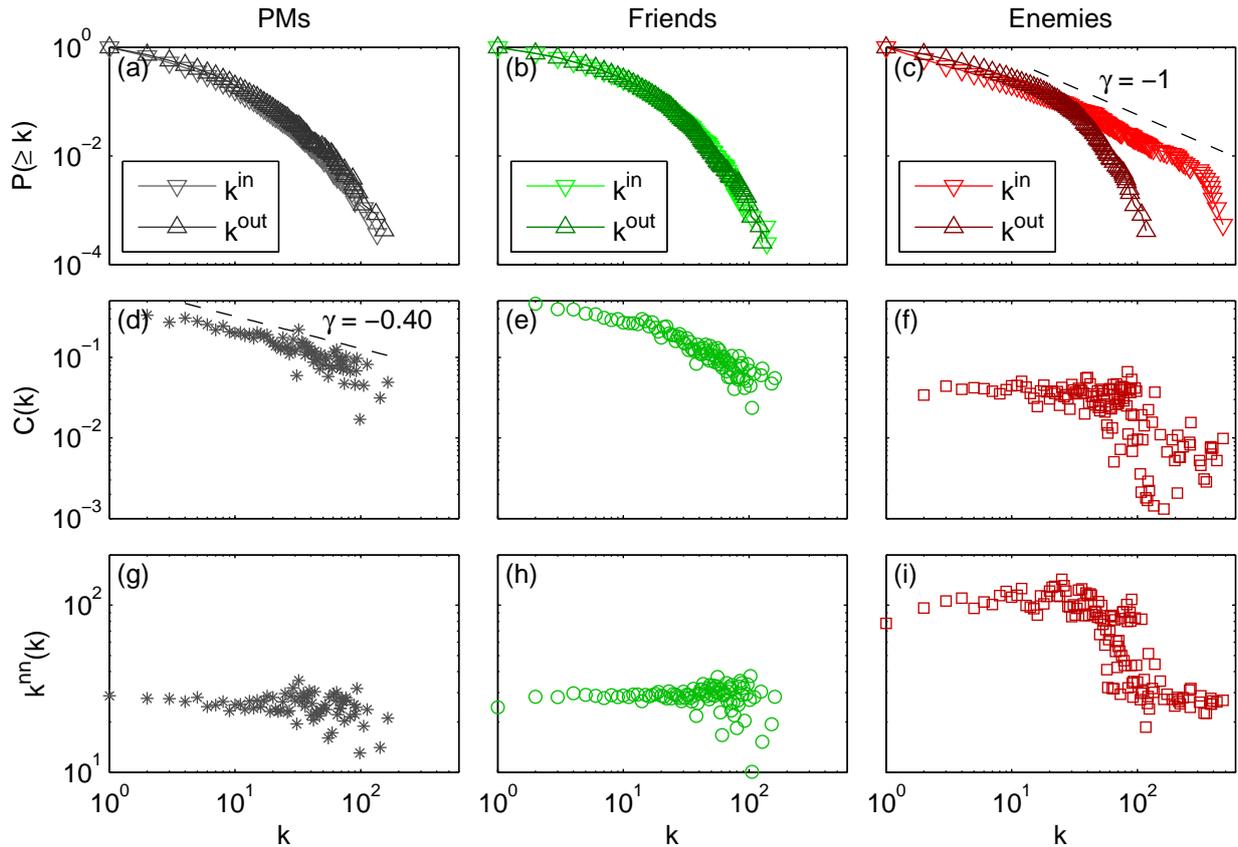}    
    \end{center}
    \caption{Cumulative degree distribution of (a) PM, (b) friend and (c) enemy networks; clustering coefficient $C$ as a function of  degree for the (d) PM, (e) friend and  (f) enemy networks; nearest neighbor degree $k^{\mathrm{nn}}$ versus degree of  the (g) PM, (h) friend, and (i) enemy networks.
    All distributions were taken at the last day.}
    \label{fig:pkckknn}
\end{figure*}

\subsection{Measurement of triad significance profiles}
\begin{figure}[!t]
    \begin{center}
        \includegraphics{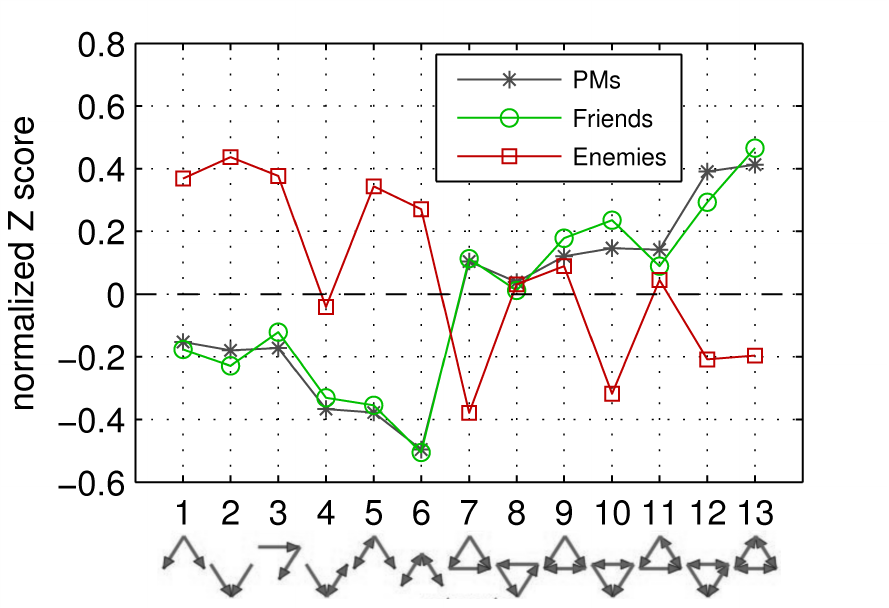}
    \end{center}
    \caption{Triad significance profiles for the three network types (day 445).
    }
    \label{fig:tspsingle}
\end{figure}

For drawing random networks from the $U(X_{*+}, X_{+*}, M^{*})$ distribution we use the same switching algorithm and Monte Carlo method as \cite{milo2004sea,milo2002nms}. We use the same program \verb|mfinder|\footnote{We used version 1.2. \\
\href{http://www.weizmann.ac.il/mcb/UriAlon}{\ttfamily{http://www.weizmann.ac.il/mcb/UriAlon}}.}. 
See \cite{roberts2000sms,milo2003ugr} for details. We calculate the TSPs for all three network types at the last day with the following parameters: 100 random networks, each generated by performing $Q \cdot L$ switches, where $Q$ is drawn uniformly from $\{100, \ldots, 200\}$. Resulting TSPs are displayed in fig.~\ref{fig:tspsingle}.

\subsection{Time evolution of TSPs}
\begin{figure}[!t]
    \begin{center}
        \includegraphics{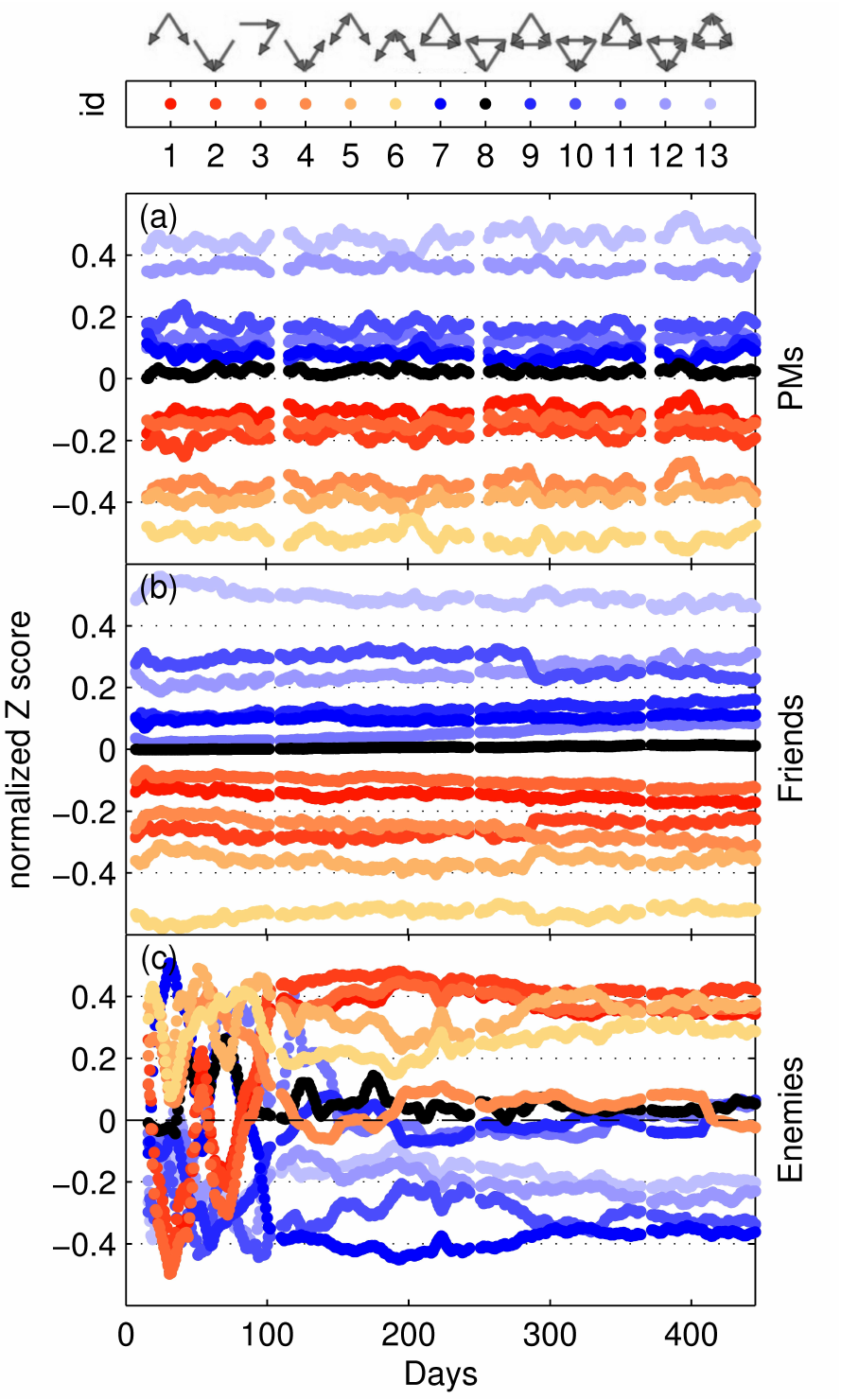}
    \end{center}
    \caption{Evolution of triad significance profiles for (a) PM, (b) friend, and (c) enemy networks. A moving average filter with a time window of 7 days was used for smoothing.}
    \label{fig:tspevolution}
\end{figure}

Figure \ref{fig:tspevolution} shows the time evolution of TSPs in PM, friend, and enemy networks, each day measured with the same parameters as in the previous section. For visual clarity, all single $Z$ score evolutions were smoothed with a moving average filter using a time window of 7 days. Z scores of PM networks stay constant, see fig.~\ref{fig:tspevolution} (a). For friend networks the order of $Z$ scores stays relatively constant except for the pairs $\{2, 4\}$ and $\{10, 12\}$, which switch order abruptly at day $\approx\,290$, 
 fig.~\ref{fig:tspevolution} (b).  At this time, a new game feature was introduced in \verb|Pardus|, which allowed players to join \emph{syndicates}. Because of this a number of players reconsidered their friend and enemy relations. Besides these abrupt changes, some trends are discernable, such as the slow decrease of triad classes $1$ and $3$ or the increase of triad classes $9$ and $11$. In enemy networks, fig.~\ref{fig:tspevolution} (c), TSPs undergo heavy fluctuations but settle into three groups of triad classes after some hundred days: $\{1, 2, 3, 5, 6\}$ being highly overrepresented ($Z$ score $>0.2$), $\{4, 8, 9, 11\}$ being neither clearly over- nor underrepresented ($Z$ score between $-0.1$ and $0.1$), and $\{7, 10 , 12, 13\}$ being highly underrepresented  ($Z$ score $<-0.15$).  

\subsection{Measurement of triad transitions}
To obtain insight in {\em triad dynamics}, we directly count all transformations of all triads in friend networks, enabling us to deduce empirical transition probabilities from one type to another. We measure $13 \times 16$ matrices $\mathbf{\Pi}^{d}$ of day-$d$-to-day-$(d+50)$ triad transition probabilities for each day $d \in \{150,\ldots,200\}$. This matrix is constructed by using an algorithm adapted from \cite{batagelj2001stc}. Its entries $\pi^{d}_{ij}$ are the empirical probabilities that triads of class $i$ on day $d$ become triads of class $j$ on day $d+50$. Note that the sum of each row of $\mathbf{\Pi}^{d}$ equals $1$. Values between differing rows are not directly comparable due to a highly heterogeneous triad census. For example, the census of all connected triads of day 200 reads (20503, 19872, 13286, 36737, 48320, 58137, 1510, 20, 1209, 4862, 453, 4788, 7887), ordered by increasing triad id. The number of unconnected triads, especially null triads, is much larger namely $\binom{N}{3}$ minus the number of connected triads; in our case this amounts to an order of magnitude of $\sim10^{10}$. Similarly, we define the matrices $\mathbf{K}^{d}$ containing empirical transition counts $k^{d}_{ij}$ of triads of class $i$ becoming class $j$.  For both matrices only those triads are counted in which all three of the involved characters still exist on day $d+50$.

We denote the matrix of element-wise time averages of $\mathbf{\Pi}^{d}$ and $\mathbf{K}^{d}$ over all considered days by $\mathbf{\Pi}$ and $\mathbf{K}$, respectively.  Entries $k_{ij}$ of matrix $\mathbf{K}$ are empirical average $50$-day transition counts of triads changing from class $i$ to class $j$, entries $\pi_{ij}$ of matrix $\mathbf{\Pi}$ the corresponding transition probabilities. Figures \ref{fig:pifriends} (a) and (b) show $\mathbf{\Pi}$ and $\mathbf{K}$. Figure \ref{fig:pifriends} (c) displays the matrix of asymmetries in $50$-day transition counts $\mathbf{K}$ for friend networks, i.e. $\mathbf{K}-\mathbf{K}^{T}$. Similar matrices for enemy networks have very different entries (not shown), matrices for PM networks were not calculated.


\section{Discussion}
\label{sec:discussion}
We proceed by discussing the results of section \ref{sec:results} and bring them into perspective.

\subsection{Preferential attachment only in enemy networks}
\label{sec:pa}
The model of preferential attachment assumes that nodes which link to a network for the first time preferably  attach to nodes with high degrees \citep{barabasi1999esr}. If preferential attachment holds in its classical form, the following facts should be observed:
\begin{enumerate}
	\item
	  Linking probability $P(k) \propto k^{\alpha}$, with $\alpha = 1$
	\item
	  Degree distribution follows a power-law $P(k) \sim k^{-\gamma}$
	\item
	  Clustering coefficient $C$ versus degree $k$ is uniform
\end{enumerate}

In our data all three points do \emph{not} hold in friend networks, figs.~\ref{fig:pa}, \ref{fig:pkckknn} (b) and (e). The exponents derived from linking probability versus degree is $\alpha \approx 0.62$ and definitely not $\alpha = 1$. Note the existence of  `super-preferentiality' for very popular characters -- a similar effect has been measured in \cite{leskovec2008mes} for \verb|LinkedIn|, a social networking site for professional contacts.  There an exponent  $\alpha = 0.6$ is reported. The degree distributions in friend networks  do not follow a power-law, the clustering coefficient versus degree exhibits a clear downward trend. Note the clearly negative value of $r_{C(k)}$ at the last days, fig.~\ref{fig:networkproperties} (k). The exponent of the clustering coefficient $C(k)$ versus degree $k$ is  $\gamma \approx -0.4$. About the same exponent has been measured in a game-theoretic model on co-evolving networks \cite{biely2005psd}; an exponent of $\gamma \approx -0.33$ was found in another large-scale social network \citep{csanyi2004sls}.

For enemy networks the situation is different, see figs.~\ref{fig:pa}, \ref{fig:pkckknn} (c) and (f). The exponent from linking probability versus degree $\alpha \approx 0.90$ is closer to $\alpha = 1$, the distribution of in-degrees follows an approximate a power-law with exponent $\gamma \approx 1$ (in the cumulative distribution). Clustering coefficients and degrees are to a large extent independent; deviations for large degrees can be explained by two different mechanics of marking enemies, see section \ref{sec:commonenemies}. Note that -- while not clearly visible on the last day's plot (fig.~\ref{fig:pkckknn} (c)) -- we find the distribution of out-degrees in enemy networks separated into two regimes following approximate power-laws  with different exponents,  $\gamma \approx -0.6$ and $\gamma \approx -2.5$, respectively (in the cumulative distribution). This is the case for many days in the Artemis universe and also for most days in the Orion universe (not shown). Actual preferential attachment has been measured in relatively few works \citep{jeong2003mpa,leskovec2008mes,hu2009dmo}; the evolution of preferential attachment parameters was measured in \cite{csardi2007mik}.

In conclusion the model of preferential attachment can not be applied for friend networks.  The situation of the  enemy networks might be closer to a PA mechanism, but also there the situation is more intricate.

\subsection{Confirmation of the Weak ties hypothesis}
The Weak ties hypothesis of Granovetter is an important proposition of sociology and builds upon the assumption that ``the degree of overlap of two individual's friendship networks varies directly with the strength of their tie to one another'' \citep{granovetter1973swt}. By Granovetter's paradoxical formulation of the Weak ties hypothesis (``The Strength of Weak Ties''), weak ties (e.g. casual acquaintanceships) are proposed to be strong in the sense that they link communities in an essential way -- i.e. they are {\em local bridges} of high degree -- while strong ties (standing for e.g. good friendships) correspond to replacable intra-community connections. Under given social balance assumptions, except for very unlikely conditions, ``\emph{no strong tie is a bridge}'', and ``all bridges are weak ties'' \citep{granovetter1973swt}. As an intuitive notion of \emph{strength} of an interpersonal tie, Granovetter mentions ``the amount of time, the emotional intensity, the intimacy (mutual confiding), and the reciprocal services which characterize the tie''.

Quantitatively  the hypothesis concerning the connection between tie strength and overlap of friendship circles should manifest itself in an increasing function of overlap,  $O(w)$ , versus weight. This is clearly the case  for PM networks, fig.~\ref{fig:obow} (b), where an approximate  cube root law is suggested, 
\begin{equation} 
O(w) = w^{0.30} \approx \,  ^3\!\!\! \sqrt{w}. 
\end{equation}
Note that our method does not encounter  sampling issues as in e.g. \cite{onnela2007als}; in this sense our results are free of bias.

A direct way of testing the Weak ties hypothesis is to examine the correlation between betweenness and overlap. By the Weak ties hypothesis, the overlap $O(b)$ as a function of betweenness should be decreasing. Our data strongly confirms this prediction, fig.~\ref{fig:obow} (a) and suggests an inverse square root law. Logarithmically binned average values lie on a line with slope $\gamma \approx -0.54$.
\begin{equation} 
O(b) = b^{-0.54} \approx \frac{1}{ \sqrt{b}}. 
\end{equation}
These results are in agreement with mobile phone call network data  \cite{onnela2007als},  and are robust across game universes and accumulation times. The Weak ties hypothesis has also previously been tested by \cite{friedkin1980tsf} on a small-scale social network of biologists.

\subsection{Confirmation of triadic closure}
\subsubsection*{Triad significance profiles (implicit evidence)}
\begin{figure}[!ht]
    \begin{center}
        \input{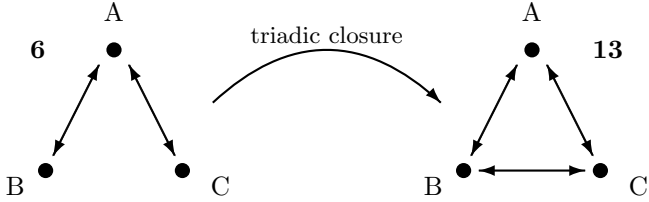}
    \end{center}
    \caption{One transition representing triadic closure:  id 6 $\to $ id 13.}
    \label{fig:transition}
\end{figure}

Comparing TSPs of various types of networks can reveal ``superfamilies'' of evolved or designed networks which have similar local structures in common \citep{milo2004sea}. In the following we use the TSP to confirm another important prediction of \cite{granovetter1973swt}. This conjecture follows balance considerations of \cite{heider1946aac} and reads as follows: In a social network in which there exist weak and strong (or no) ties between individuals, ``the triad which is most \emph{unlikely} to occur, [\ldots] is that in which A and B are strongly linked, A has a strong tie to some friend C, but the tie between C and B is absent'' \citep{granovetter1973swt}. The phenomenon of \emph{triadic closure} \citep{rapoport1953sit} states that individuals are driven to reduce this cognitive dissonance, fig.~\ref{fig:transition}. Because of this the triad in which there exist strong ties between all three subjects A, B and C should appear in a higher than expected frequency.

Following the considerations of Granovetter about tie strength we identify the concept of weak/strong ties with asymmetric/mutual dyads in our digraphs. Translated into our formalism the hypothesis reads: ``In friend networks, triad class 6 should have smallest $Z$ score, triad class 13 should have highest $Z$ score''. In other words, triad class 13 should be the network's strongest three-node \emph{motif}, triad class 6 should be it's strongest three-node \emph{antimotif} \citep{milo2002nms}. More generally, if we focus on completeness, we expect underrepresentation of the incomplete triad classes 1--6 and overrepresentation of the complete triad classes 7--13. Note that quantitative evidence for triadic closure given this way is implicit at best, since the overrepresentation (underrepresentation) of triad class 13 (6) does not explain \emph{how} or \emph{if} there is a direct connection in the evolution of these triad classes. We give explicit evidence for triadic closure in the next section by measuring triad transition dynamics.

\begin{figure}
    \begin{center}
        \includegraphics[width=0.49\textwidth]{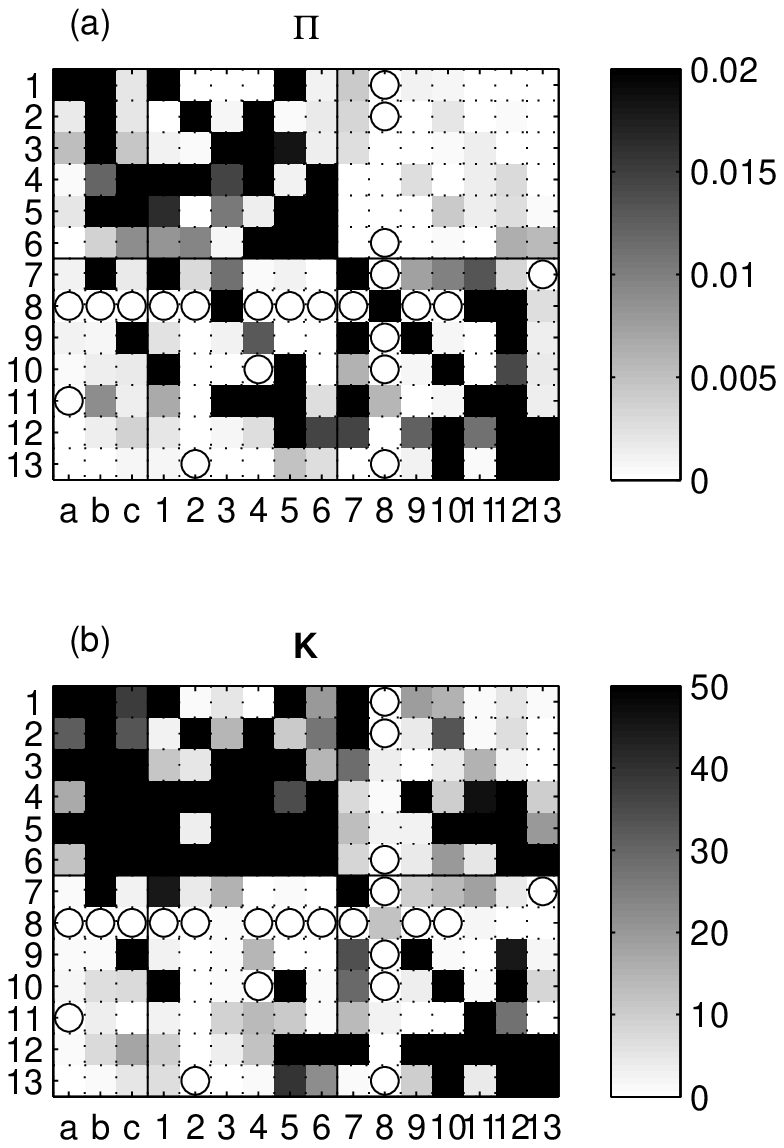}
        \includegraphics[width=0.49\textwidth]{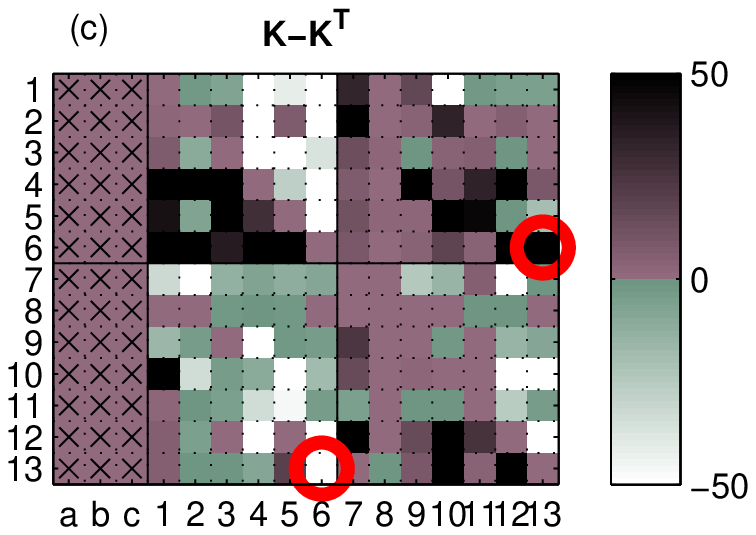}
    \end{center}
    \caption{(a) Matrix $\mathbf{\Pi}$ of empirical average $50$-day transition probabilities of triad classes in friend networks over the days 150 to 200. Circles mark transitions which never occured. Black squares mark average transition probabilities $\geq 0.02$. (b) Matrix $\mathbf{K}$ of empirical average $50$-day transition counts of triad classes in friend networks. Black squares mark average transition counts $\geq 50$. (c) Matrix of asymmetries between empirical average $50$-day transition counts $\mathbf{K}$ of triad classes in friend networks. Black and white squares mark average transition counts with asymmetries $\geq 50$. Black crosses mark entries without data. Red circles mark the asymmetry of transitions between triad classes $6$ and $13$.}
    \label{fig:pifriends}
\end{figure}

The question about the reverse situation, networks of negative ties, has been raised in the outlook of \cite{granovetter1973swt} but, to our knowledge, has never been measured on large scales. Following the same social balance considerations we expect reversed roles of completeness: Instead of the \emph{absence} of a completing third link, its \emph{presence} should cause cognitive dissonance (however, note that a complete triad with only negative links may be seen as ambiguous concerning balance \citep{doreian2004ehs}). Thus triad classes 1--6 should be overrepresented, triad classes 7--13 underrepresented in enemy networks.

For  friend and PM networks, excellent agreement is found with Granovetter's prediction: Triad class 6 has the minimum, class 13 a maximal $Z$ score, fig.~\ref{fig:tspsingle}. Our findings further coincide with the TSPs of the superfamily of social and hyperlink networks found in \cite{milo2004sea}  and with TSPs of other social networks \cite{hamasaki2009nae}.  Concerning enemy networks, we observe confirmation of our reverse hypothesis to a large extent: Most enemy $Z$ scores have opposite signs of those in friend networks. Note the exceptions: triad id 4 is not clearly overrepresented, ids 9 and 11 are not clearly underrepresented. The circular triad (id 8) should be considered an exceptional or 'neutral' class, having no clear tendency in all network types.

In the above paragraphs triad significance profiles of one day were analyzed.  By measuring evolutions of TSP, we are able to confirm the robustness of the results. As one can see in fig.~\ref{fig:tspevolution} (c), TSPs may need some hundred days to reach an approximate steady state. On the other hand it is apparent from fig.~\ref{fig:tspevolution} (b) that as social networks evolve, their microscopic structures do not always stay completely constant. Sudden jumps in the TSP  trajectories signal abrupt global systemic changes.

\subsubsection*{Triad transition rates (explicit evidence)}
We first focus on transitions between the groups of incomplete and complete connected triads, i.e. on values in the center right and lower center zones of $\mathbf{K}$. According to the hypothesis of triadic closure, $k_{6,13}$ should contain high values, its counterpart $k_{13,6}$ lower ones. As apparent from fig.~\ref{fig:pifriends} (b), this is the case: $k_{6,13} = 305.5 > 22.6 = k_{13,6}$. This result is directly visualized in fig.~\ref{fig:pifriends} (c), which depicts the matrix $\mathbf{K}-\mathbf{K}^T$. The darker a square $ij$, the higher the outflow $i \rightarrow j$ compared to the inflow $j \rightarrow i$; the lighter a square $ij$, the higher the inflow $j \rightarrow i$ compared to the outflow $i \rightarrow j$.

In general, we measure more incomplete $\rightarrow$ complete transitions between connected triad classes (upper right zone of $\mathbf{K}$) than vice versa (lower center zone of $\mathbf{K}$). On the other hand, some exceptions can be identified, for example $k_{5,13} = 20.0 < 39.2 = k_{13,5}$. Whenever these exceptions appear they are comparably  mild; again, observations are robust across game universes and time spans.

\subsection{Network densification}
In \cite{leskovec2007ged} intriguing observations concerning universal features of growing real-world networks have been made. Their empirical observations apparently challenge two conventional assumptions of popular network models such as PA \citep{barabasi1999esr}, namely  constant average degrees and slowly growing network diameters:
\begin{enumerate}
        \item \emph{Shrinking diameters:} As networks grow their diameters decrease.
        
	\item \emph{Densification power-laws:} Over time, networks become more dense. Densification -- as measured as number of edges versus number of nodes -- follow a power-law. The average degree grows. 
	\end{enumerate}

All growing networks measured in this work confirm the observations of growing average degrees and shrinking diameters (fig.~\ref{fig:networkproperties} (g) shows shrinking geodesics; we observe the same evolution for diameters and effective diameters as defined in \cite{leskovec2007ged} (not shown)). Concerning power-law densification, 
even though our data does not allow for a statistically conclusive quantitative statement, visual inspection clearly reveals that growth is super-linear, fig.~\ref{fig:linksversusnodes}. Network densification has previously  been studied under the name of  \emph{accelerated growth} \citep{dorogovtsev2001eag, dorogovtsev2003enb}. Growing average degrees were observed in all three time evolution studies of growing networks we are aware of \citep{holme2004sat,leskovec2007ged,ravasz2004eha}.

From decreasing distances it follows naturally that global efficiency increases, fig.~\ref{fig:networkproperties} (i). At the same time evolution of local efficiency follows the evolution of clustering coefficients, fig.~\ref{fig:networkproperties} (f). 
PM networks have approximately the same local and global efficiency (around $0.25$), in friend networks local efficiency (around $0.33$) is higher than global efficiency (around $0.22$), in enemy networks local efficiency is much lower (around $0.04$) than global efficiency (around $0.28$).

\subsection{Inconclusive social balance dynamics}
Social balance theory goes back to the cognitive balance considerations of \cite{heider1946aac}. A complete triad $n_i n_j n_k$ is defined to be \emph{balanced} if the product of signs $s_{ij}s_{jk}s_{ki} = 1$, and is \emph{unbalanced} otherwise. Members of a balanced complete triad thus fulfill the following adage \citep{antal2006sbn,heider1946aac}:

	-- a friend of my friend is my friend
         
	-- a friend of my enemy is my enemy
	
	-- an enemy of my friend if my enemy 
	
	-- an enemy of my enemy is my friend 
	
In physics the first statement corresponds to a ferromagnetic system, the other three to a `frustrated' system. In graph theory, the concept of social, or structural, balance has been generalized to an arbitrary amount of subjects by \cite{cartwright1956sbg}; hypotheses about the evolution of social balance have been conjectured \citep{doreian2004ehs, doreian2009pss}. We  measure the evolution of social balance by using optimizational partition algorithms implemented in \verb|Pajek|\footnote{We use version 1.24. \citep{doreian1996pas,denooy2005esn,doreian2009pss}}. 
 Here we face  three concrete problems:
(i) \emph{Algorithmic complexity:} Due to algorithmic complexity runtime diverges for large numbers of nodes (\verb|Pajek| limits the number of nodes to 250). Thus for measurements of balance we are forced to select groups of characters.
(ii) \emph{Group selection:} Characters have different sign-up dates. This starts to matter for  any selected  group when long-time considerations are carried out. One gets  inhomogeneous groups, where some characters have a long history of relations whereas  others have not.
(iii) \emph{Growth of average degrees:} The permanent growth of average degrees, fig. \ref{fig:networkproperties} (a), is inconsistent with the necessary assumption of constant degrees, such as taken as basis for the monastery study of \cite{sampson1968npc} analyzed in \cite{doreian1996pas}. This assumption is also needed in models and analytical work \citep{antal2006sbn}, where dynamics takes place only on complete graphs, i.e. graphs displaying a dichotomy of link types (positive or negative link) in contrast to  the trichotomy  (positive, negative or no link) of the \verb|Pardus| networks. 

Ignoring these three issues, our attempts to measure the evolution of social balance in several groups of characters at various time scales and cluster sizes yield no conclusive results. Neither an increase nor a decrease in balance could be observed. Measuring social balance in our social system is inherently futile, for several further reasons: changes of links between neighbors of neighbors may take a lot more time to notice for people than classically assumed. Also, system complexity is so high (factions, alliances, wars, \ldots), that simple analytical steady states (e.g. two internally positively --  but among each other negatively connected sets of nodes) are impossible to reach.  Compared to the slow propagation of link information there seems to be too much noise in the system -- time scales are separated to a too great extent. For a deeper discussion  about these and several other problems related to social balance theory and inconsistencies in empirical findings see \cite{hummon2003sds, doreian2004ehs}.

Note that the friend relation is transitive, but not so the enemy relation.
\footnote{A binary relation $R$ is transitive if $xRy$ and $yRz$ implies $xRz$.} By construction, complete triads in the reflexive closure of  friend (enemy) networks are balanced (unbalanced).  In enemy networks one therefore expects less complete triads and  clustering coefficients $C$ closer to their random graph values -- i.e. smaller $C/C_{\mathrm{r}}$ -- than in networks of friends. As apparent from fig. \ref{fig:networkproperties}  (e), this is fully confirmed.

\subsection{Confirmation of the Dunbar number}
Out-degrees of networks studied here are limited by $k^{\mathrm{out}} \approx 150$, fig.~\ref{fig:pkckknn}  (a), (b), and (c).
This number was conjectured by \cite{dunbar1993cen} to be a limiting number in group sizes of humans and human-like mammals, due to their limited cognitive capacities. 

\subsection{Two categories of enemies}
\label{sec:commonenemies}
It has been suggested that identifying negative social tie mechanics is more important for gaining insight in social group dynamics than identifying mechanics of positive social ties \citep{labianca2006esl}. Our measurements provide first steps toward this direction. 

We observe that in-degrees in enemy networks grow much bigger ($k^{\mathrm{in}} \approx 500$) than in-degrees in friend and PM networks ($k^{\mathrm{in}} \approx 150$), figs.~\ref{fig:pkckknn} (b) and (c). Also, the assortativity coefficient $r_{\mathrm{undir}}$ of enemy networks is clearly negative, fig.~\ref{fig:networkproperties} (l). In the average neighbor degree $k^{\mathrm{nn}}$ versus degree $k$ one observes two classes or types, fig.~\ref{fig:pkckknn} (i). The first class consists of characters with low degrees ($k < 50$), all having an average neighbor degree of  $k^{\mathrm{nn}} \approx 100$. The second class of characters has a degree of $k > 100$ and an average neighbor degree of  $k^{\mathrm{nn}}
\approx 28$. Between these classes a sharp transition occurs. This impression is robust across time and game universes.

As mentioned in section \ref{sec:pa},  on many individual days the distribution of out-degrees of enemy networks is separated into two regimes roughly following power-laws with markedly different exponents. We find clear qualitative differences between in- and out-degree distributions in enemy networks, in contrast to the other network types. Further, reciprocity is very low in enemy networks, as opposed to high reciprocity in the other network types, fig.~\ref{fig:networkproperties} (j).
These measurements suggest two distinct mechanics of enemy marking dynamics at work:
\begin{enumerate}
	\item \emph{Private enemies:} A player who directly experiences a negative (asocial) action by another player, such as an attack of her building or a verbal insult, is likely to \emph{immediately} react by marking the offender as enemy. If the two involved players keep this a private affair, only a local, dyadic vendetta without envolvement of more players may ensue.
	\item \emph{Public enemies:} Some players have a destructive personality or, more commonly, want to \emph{try out} a destructive personality \citep{castronova2005swb}. For this purpose they may (role)play evil characters, such as `pirates'. These players tend to take enjoyment in destroying other players' work or see it as their `task'. Anonymity of the internet facilitates this behavior to some degree since possible social repercussions in real-world reputation are absent. For this reason these few individuals tend to cause a lot of offenses to a big number of players. If such a subject is identified by the community (players are very busy in using the forums to keep others up-to-date of the latest offenses), she may receive pre-emptive enemy markings, either by friends of offended friends or by otherwise non-involved players who happen to read the forums. This destructive behavior and the indirect marking mechanism leads to the emergence of `public enemies', i.e. a few characters with a very high in-degree of enemy markings. The strength of positive social ties is likely to be boosted by people who share the same common enemies: ``A world that includes self-proclaimed and loudly advertised Evil people running about represents a great boon to those who are hungry to fight for the Good. Without Evil people, who could be Good?'' \citep{castronova2005swb}. 
\end{enumerate}

It is an open question to which extent negative social behavior in real society deviates from the behavior of humans in our game society.  Due to lack of other  high-frequency analyses on large-scale negative tie networks,  it remains to be established whether the above findings can be referred to as `universal'.

\subsection{Differences in network types}
Different network types have different properties \citep{newman2003wsn, milo2004sea}, and show a different evolution of these properties. The clear asymmetries of evolutions of reciprocity, clustering per degree, and assortativity between friend and enemy networks are obvious, fig.~\ref{fig:networkproperties} (j), (k), and (l). These asymmetries originate from differences in the corresponding formation processes. Intuitively, someone either is your enemy or not -- which is apparent to determine and declare --, but friendship comes in several shades of gray more likely being dependent on long-time social dynamics. This intuition is confirmed by \cite{labianca2006esl}: ``The evolution of negative relationships may be very different from positive relationships. Friendship development is viewed as a gradual process. According to social penetration theory [\ldots],  friendship development proceeds from superficial interaction in narrow areas of exchange to increasingly deeper interaction in broader areas. [\ldots]  Qualitative work indicates that negative relationship development is a much faster process that tends to lead to the other person being included in coarse-grained categories such as ``rival'' or ``enemy''. Thus, the formation of negative relationships is not the mere opposite of the way that positive relationships form.''

Further, we observe that reciprocities as well as assortative mixing coefficients need much more time to reach steady state in friend networks than in other networks. It seems to be clear why equilibria of the properties of PM networks are reached fast: Writing or responding to a PM is an administrative, immediately executable action to a large extent independent of social ties. On the other hand, choosing your friends or the reciprocation of friendships is a much more delicate operation, necessitating in-depth considerations about e.g. social balance.

Another issue is the possible `import' of social ties. In a recent study of online social networks, the effect of a transition from assortative social networks to disassortative or non-assortative networks has been observed  \citep{hu2009dmo}. While there it could not be checked if networks are at their initial stage or not, here we are able to do so (for the Artemis and Pegasus universes). Our following hypothesis coincides with the reasoning of \cite{hu2009dmo}: At the initial stage of social online networks, some people may `import'  their social relations from existing previous ones. This is possible since many players who previously played in Orion created characters in Artemis and Pegasus when these new universes opened. At later times (in our case 100 or more days), evolving new relations of online interaction take overhand and decrease assortativity, down to disassortativity. The same effect may lead to the observed changes in  clustering coefficients, reciprocity, and  other network properties.


\section{Conclusion}
\label{sec:conclusion}
We explore novel possibilities of a quantification of human group-behavior on a fully empirical and  falsifiable basis. We study network structure and its evolution of several social networks extracted from a massive multiplayer online game dataset. Practically all actions of all 300,000 players over a period of three years are available within one  unique and coherent source. Players live a second economic life and are typically engaged  in a multitude of social activities within the game. With this data we can show for the first time marked differences in the dynamics of friend and enemy dynamics. A detailed  analysis of high-frequency log files focuses on three types of social networks and allows to subject a series of long-standing social-dynamics hypotheses to empirical tests with extraordinary precision. Along these lines we propose two social laws in communication networks, the first expressing betweenness centrality  as the inverse square of the overlap, the second relating communication strength to the cube of the overlap. These laws not only provide strong quantitative evidence for the validity of the Weak ties hypothesis of Granovetter, they are also fully falsifiable. Our study of triad significance profiles confirms several well-established assertions from social balance theory. We find  overrepresentation (underrepresentation) of complete (incomplete) triads in networks of positive ties, and vice versa for networks of negative ties. We measure empirical transition probabilities between triad classes and find  evidence for triadic closure, again with unprecedented precision. We compare our findings with data from non-virtual human groups and conclude that online game communities should be able to serve as a model for a wide class of human societies. We demonstrate the realistic chance of establishing socio-economic laboratories which allow to measure dynamics of our kind at levels of precision so far only known from the natural sciences.


\section*{Acknowledgements}
\label{sec:acknowledgements}
We are indebted to Werner Bayer for compiling \verb|Pardus| backup data and for providing computer power for TSP calculations. 
This work was supported in part by Austrian Science Fund FWF P 19132. 


\bibliographystyle{elsarticle-harv}
\bibliography{bibszellthurner2009}

\end{document}